\documentclass[journal]{IEEEtran}
\usepackage{graphicx}
\usepackage{amsmath,amssymb,dsfont,bbm,epstopdf,pgfplots,mathtools,enumitem,mathrsfs, bbm, algpseudocode , graphicx, dblfloatfix, booktabs, physics, algorithm,algcompatible, derivative, physics} 
\usepackage[normalem]{ulem}
\usepackage{color}
\usepackage{cite}
\usepackage{graphics}
\usepackage{tikz}
\usepackage{pgfplots}
\usepackage{subcaption}
\usepackage[utf8]{inputenc} 
\usepackage[T1]{fontenc}
\usepackage[left=0.6in,right=0.6in,top=0.71in, bottom = 0.94in]{geometry}
\usetikzlibrary{shapes, arrows, decorations.markings, arrows.meta}
\usetikzlibrary{patterns} 
\allowdisplaybreaks
\graphicspath{{.}{./Figures/}} 
\allowdisplaybreaks[4] 

\newtheorem{lemma}{Lemma}

\newtheorem{theorem}{Theorem}
\newtheorem{proposition}{Proposition}
\newtheorem{remark}{Remark}
\newtheorem{definition}{Definition}

\newcommand{\R}{{\mathbb R}}

\renewcommand{\Pr}{{\mathbb{P}}}

\newcommand{\vect}[1]{\boldsymbol{#1}}
\newcommand{\vectsf}[1]{\boldsymbol {\mathsf {#1}}}

\definecolor{ForestGreen}{rgb}{0.0, 0.5, 0.0}

\renewcommand{\P}{\mathsf{P}}
\newcommand{\D}{\mathsf{D}}
\newcommand{\C}{\mathsf{C}}
\renewcommand{\R}{\mathsf{R}}
\renewcommand{\P}{\mathsf{P}}

\newcommand{\U}{\mathsf{U}}
\newcommand{\e}{\mathsf{e}}
\newcommand{\s}{\mathsf{s}}
\newcommand{\bu}{\beta_{\U}}

\newcommand{\pli}{p_{\ell_i}}

\newcommand{\Bdet}{\mathcal B_{\text{detect}}}
\newcommand{\Barr}{\mathcal B_{\text{URLLC}}}
\newcommand{\Bdec}{\mathcal B_{\text{decode}}}

\newcommand{\Bdt}{B_{\text{dt}}}
\newcommand{\Bdc}{B_{\text{dc}}}
\newcommand{\vbj}{\vect V_{b,j}}
\newcommand{\ybcj}{\vect Y_{b,c,j}}
\newcommand{\sbjs}{\vect S_{b,j}^{(2)}}
\newcommand{\sbjf}{\vect S_{b,j}^{(1)}}
\newcommand{\xbjs}{\vect X_{b,j}^{(\s,2)}}
\newcommand{\xbjsf}{\vect X_{b,j}^{(\s,1)}}
\newcommand{\nbcj}{\vect N_{b,c,j}}
\newcommand{\tepf}{\tilde \epsilon_{\U,1}}
\newcommand{\teps}{\tilde \epsilon_{\U,2}}

\newcommand{\I}{\vectsf{I}}
\newcommand{\bc}{\beta_u}
\newcommand{\bsf}{\beta_{s,1}}
\newcommand{\bss}{\beta_{s,2}}
\newcommand{\alc}{\alpha_u}
\newcommand{\alsf}{\alpha_{s,1}}
\newcommand{\alss}{\alpha_{s,2}}
\newcommand{\sigbj}{\sigma^2_{b,j}}
\newcommand{\sigyvj}{\sigma^2_{y|v,j}}
\newcommand{\sigy}{\sigma^2_{y,j}}
\newcommand{\sigysfj}{\sigma^2_{y|s^{(1)},j}}
\newcommand{\sigysvj}{\sigma^2_{y|s^{(2)}v,j}}
\newcommand{\sigyssj}{\sigma^2_{y|s^{(2)},j}}
\begin{document}
\title{ An Integrated Sensing and Communication System for Time-Sensitive Targets with Random Arrivals}

\author{\IEEEauthorblockN{Homa Nikbakht, Yonina C.~Eldar, \IEEEmembership{Fellow,~IEEE},  and H.~Vincent Poor, \IEEEmembership{Life Fellow,~IEEE }}

	\thanks{
	Part of this work is accepted for presentation at the IEEE International Symposium on Information Theory (ISIT), 2025 \cite{HomaISIT2025}.  
}
}
\maketitle
\begin{abstract}

In 6G networks,  integrated sensing and communication (ISAC) is envisioned as a key technology that enables wireless systems to perform joint sensing and communication using shared hardware, antenna(s) and spectrum. ISAC designs facilitate emerging  applications such as digital twins, smart cities and autonomous driving.  Such applications also demand ultra-reliable and low-latency communication (URLLC), a feature that was first introduced in 5G and is expected to be further enhanced in 6G.  Thus, an ISAC-enabled URLLC system can prioritize critical and time-sensitive targets and ensure information delivery under strict latency and reliability constraints. We propose a bi-static multiple-input multiple-output (MIMO) ISAC system to detect the arrival of URLLC messages and prioritize their delivery. In this system, a dual-function base station (BS) communicates with a user equipment (UE) and a sensing receiver (SR) is deployed to collect echo signals reflected from a target of interest.  The BS regularly transmits messages of enhanced mobile broadband (eMBB) services to the UE. During each eMBB transmission, if the SR senses the presence of a target of interest, it immediately triggers the transmission of an additional URLLC message. To reinforce URLLC transmissions, we propose a dirty-paper coding (DPC)-based technique that mitigates the interference of both eMBB and sensing signals. To decode the eMBB message, we consider two approaches for handling  the URLLC interference: treating interference as noise (TIN) and successive interference cancellation (SIC).   For this system, we formulate the rate-reliability-detection trade-off in the finite blocklength (FBL) regime by evaluating  the communication rate of  the eMBB transmissions, the reliability of  the URLLC transmissions and the  probability of  the target detection.  
Our numerical analysis show that  our proposed DPC-based ISAC scheme significantly outperforms power-sharing based ISAC  and traditional time-sharing schemes. In particular, it achieves higher eMBB transmission rate while satisfying both URLLC and sensing constraints. 
\end{abstract}
\section{Introduction}

Integrated sensing and communication (ISAC) is enabled by higher frequency bands, wider bandwidths and denser distributions of massive antenna arrays.  
Such integration is mutually beneficial to sensing and communication tasks \cite{Visa2024,  Li2024, YLiu2024}. On the one hand, the radio wave transmission and reflection can be used to sense the environment and thus the entire communication network can serve as a sensor. On the other hand, the capabilities of high-accuracy localization, imaging, and environment reconstruction obtained from sensing can improve communication performance. ISAC designs thus offer various use cases for autonomous driving, smart factories, and other environment-aware scenarios in 6G communication networks \cite{Liu2022, Hatami2024, Mittelbach2025, Wu2022, Niu2025}.


Many ISAC use cases also require ultra-reliable and low-latency communication (URLLC), a feature introduced in 5G and is expected to evolve further in 6G  \cite{Mahmood2023, Qin2025, Luat2025, Pourkabirian2024}. For example, in autonomous driving,  ISAC plays a crucial role in detecting targets, including pedestrians and vehicles and delivering sensing information to users. URLLC is essential for ensuring the timely delivery of critical road safety information \cite{Kurma2024}. In particular, URLLC ensures 99.99\% reliability at a maximum end-to-end delay of no more than one millisecond, which is essential for many 5G and 6G applications, including industrial automation, intelligent transportation, and telemedicine \cite{Singh2024, Ding2022, Behdad2024, Zhao2022, Keshtiarast2025}.


Despite progress in achieving the required latency and reliability, 5G URLLC still does not meet all key performance metrics needed for diverse mission-critical applications. One of the main challenges arises from the random generation nature of URLLC services, as their generation is often linked to the occurrence of critical, time-sensitive events in the environment. Consequently, their arrival time becomes unpredictable for transmitting and receiving units. 
Another challenge is the coexistence of URLLC services with other 5G/6G services such as enhanced mobile broadband (eMBB) that depend largely on high transmission rate and are less sensitive to delay. Different coexistence strategies have been studied in the literature \cite{HomaEntropy2022, Song2019,Interdonato2023, HomaGlobecom2023, Wang2024}. For example,  \cite{Song2019} proposes a puncturing strategy (also known as time-sharing) in which  the on-going eMBB transmission stops upon the arrival of URLLC messages. The work in \cite{Interdonato2023} shows that a superposition coding strategy in which the transmitter simply sends a linear combination of eMBB and URLLC signals  while sharing the total transmit power between the two services outperforms the puncturing strategy.  A  dirty-paper coding (DPC) \cite{Costa1983, HomaISIT2022, Li2023Globecom} based joint transmission strategy is proposed in \cite{HomaGlobecom2023} which also outperforms the puncturing technique.  Theses studies either assume a deterministic model or a random model with Bernoulli distribution for the arrival of URLLC messages \cite{HomaGlobecom2023}, which have shortcomings in offering a practical model for the stochastic nature of this type of services. Therefore, they do not provide an effective model to address both challenges.

In this work, we address these challenges by proposing a bi-static MIMO ISAC-enabled URLLC system that supports the coexistence of URLLC with eMBB services while using  its own sensory data to trigger URLLC transmissions with no assumption on their arrival  distribution. 
 In the proposed system, the eMBB message arrives at the beginning of the transmission slot and its transmission lasts over the entire slot. Whereas, transmission of a URLLC message is triggered only when the SR detects the presence of a target. 
 
 To enable real-time joint sensing and communication in this system, we divide the eMBB transmission slot into smaller blocks. In each block, the base station (BS) transmits dual-function signals performing simultaneous communication and sensing tasks.  If the SR senses the presence of a target, it triggers the transmission of an additional  URLLC message over the next immediate block. Each block thus is either with \emph{no URLLC}  or  \emph{with URLLC} depending on whether a target is detected  in the previous block. 
 In blocks with no URLLC, we generate the dual function signal using DPC to precancel the interference of sensing signal from the eMBB transmission. In blocks with URLLC,  to increase the reliability of the URLLC transmission, we also propose a  DPC based method to generate the dual function signal. In this method, we first precancel the interference of  the sensing signal from the eMBB transmission, and then precancel the interference of both eMBB and sensing signals from the URLLC transmission. After each block, the UE attempts to decode a URLLC message using a threshold decoder. 
 
 The UE decodes  the eMBB message after the entire transmission slot. To handle URLLC interference on the eMBB transmission,  we consider two approaches: treating interference as noise (TIN), and successive interference cancellation (SIC). In the TIN approach, the decoding of the eMBB message depends on the detection of URLLC messages sent over all blocks of the transmission slot.  Under the SIC approach, prior to the decoding of the eMBB message, the UE first mitigates the interference from correctly decoded URLLC messages across all blocks. Therefore, successful eMBB decoding relies on both the accurate detection and correct decoding of URLLC messages in all blocks.
For this system, 
 we optimize the dual-function transmit waveform  to maximize the eMBB transmission rate, while ensuring that the URLLC decoding error probability remains below a threshold and the target detection probability exceeds a threshold across all blocks. 
 
 We compare our proposed DPC-based scheme with two baseline schemes: time-sharing and power-sharing schemes.    
Through numerical analyses we show that  our proposed DPC-based ISAC scheme outperforms the power-sharing and the time-sharing schemes by achieving higher eMBB rate while accommodating the URLLC constraints and the sensing constraints of the target detection. Our numerical analysis also  demonstrate the rate-reliability-detection  performance trade-off under both the TIN and SIC approaches. The results show that when there is a high reliability constraint on the URLLC transmissions,  the SIC approach outperforms the TIN approach by achieving higher eMBB transmission rate while satisfying the sensing constraints of the target detection. However, the performance gap between the two approaches decreases as the URLLC reliability constraint becomes less stringent. Our proposed ISAC-enabled URLLC system thus effectively addresses both the coexistence and the random nature generation challenges of URLLC services while outperforming the baseline time-sharing and power-sharing based systems.  


The rest of this paper is organized as follows. Section~\ref{sec:setup}  describes the general problem setup. Section~\ref{sec:coding} presents our coding schemes and our target detection strategy. Section~\ref{sec:main} discusses our main results.   Section~\ref{sec:conclusion} concludes the paper. Technical proofs are deferred to  appendices.

\section{Problem Setup}\label{sec:setup}
\begin{figure}[t]
\center
\includegraphics[width=0.44\textwidth]{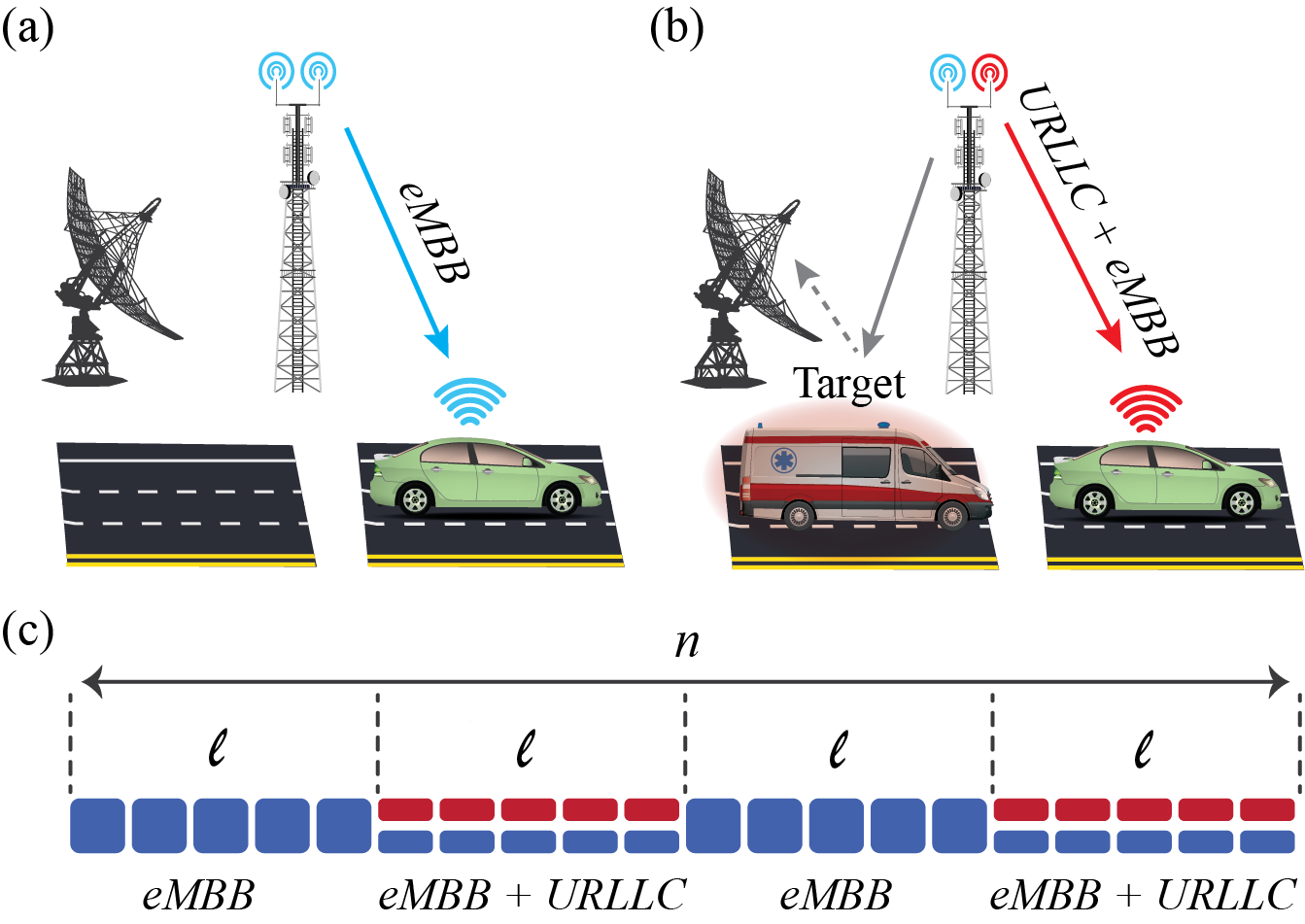}
\vspace{-0.1cm}
\caption{An illustration of the system model: (a) no target is detected, (b) detection of a target, (c) example of transmission blocks with $\eta = 4$.} 
\label{fig1}
\vspace{-0.4cm}
\end{figure}
We consider a  bi-static MIMO ISAC system where a BS is communicating with  a UE and simultaneously wishes to sense a target of interest. A SR is deployed to collect echo signals reflected from a target. We assume that the UE is not within the  SR sensing range. Hence, the SR does not receive echo signals from the UE.  The BS is equipped with $t$ transmit antennas, the UE and the SR each are equipped with $r$ receive antennas. The BS communicates both eMBB and URLLC type messages to the UE. Assume $n$ is the total communication frame length. The BS regularly transmits the eMBB message over the entire $n$ channel uses. We divide $n$ into $\eta$ blocks each of length $\ell$ (i.e., $n = \eta \cdot \ell$), as in Fig.~\ref{fig1}. The sensing task is performed across all $\eta$ blocks. In each block~$b$, if the SR senses the presence of a target, it triggers the transmission of an additional URLLC message in the next immediate block. We assume that the SR and the BS can communicate over an interference-free backhaul link. In each block~$b$, the BS thus forms its dual-function transmit signal adaptively based on the target detection outcome in the preceding block~$b-1$.
In the following sections, we explain the procedures for sensing, adaptive dual-function signal generation, and communication in such blocks. 
\subsection{Target Echo Signal Model}
The sensing task is performed across all blocks. In a given block~$b \in [\eta]$ with $[\eta]:=\{1,\ldots, \eta\}$, the SR receives the reflected echo signal $\vectsf Y_{b,s} = [\vect Y_{b,s,1} \ldots \vect Y_{b,s,r} ] \in \mathbb C^{\ell \times r}$. Upon observing $\vectsf Y_{b,s}$, the goal of the SR is to detect the presence of a target. 
The target detection problem in each block thus is formulated by defining the following two hypotheses: 
\begin{IEEEeqnarray}{rCl}
\mathcal H_0&:& \vectsf Y_{b,s} = \vectsf N_{b,s},  \\
\mathcal H_1&:& \vectsf Y_{b,s} = \vectsf X_{b}  \vectsf H_{b,s}  + \vectsf N_{b,s}, \label{eq:H1}
\end{IEEEeqnarray}
where $\vectsf H_{b,s} \in \mathbb C^{t \times r} $ is the target channel response matrix, $\vectsf N_{b,s} \in \mathbb C^{\ell \times r}$ is additive noise matrix with each entry having zero mean and unit variance, and $\vectsf X_b \in  \mathbb C^{\ell \times r}$ is the dual-function signal transmitted by the BS. We will explain shortly how this signal is formed.  The optimal detector for this problem is the likelihood ratio test (LRT) \cite{Poor} and is given by 
\begin{equation}\label{eq:test}
T_b = \log  \left ( \prod_{j = 1}^{q} \frac{f\left (\vect Y_{b,s,j}  | \mathcal H_1\right)}{f\left(\vect Y_{b,s,j} | \mathcal H_0\right)} \right) \underset{{\mathcal H_0}}{\overset{\mathcal H_{1}}{\gtrless}} \delta
\end{equation}
where   $f\left (\vect Y_{b,s,j} | \mathcal H_0\right)$ and $f\left (\vect Y_{b,s,j} | \mathcal H_1\right)$ are the probability density functions (pdf) of the observation vector under the null and alternative hypotheses, respectively, and $q:=\min \{r,t\}$. We denote the target detection and false alarm probabilities of block~$b$ by $P_{b,\D}$ and $P_{b,\text{FA}}$, respectively, which are 
\begin{subequations}
\begin{IEEEeqnarray}{rCl}
P_{b,\D} &=& \Pr [T_b > \delta | \mathcal H_1], \\
P_{b,\text{FA}} & = & \Pr [T_b > \delta | \mathcal H_0].
\end{IEEEeqnarray}
\end{subequations}
The threshold  $\delta$ is  set to ensure the desired probability of false alarm. In each block~$b$, we assume that the SR has the full knowledge of the sensing channel state information.

\subsection{Dual-Function Signal Generation}
In each block~$b\in [\eta]$,  the dual-function signal $\vectsf X_b$ is generated adaptively depending on whether a target is detected in the preceding block~$b-1$. More specifically, the BS transmits the eMBB message $m_e$ across all $\eta$ blocks, where $m_{\e}$ is uniformly drawn from the  set $\mathcal M_{\e} : = \{1, \ldots, M_{\e}\}$. 
 In each block~$b$, the BS transmits an additional URLLC message $m_{b,\U}$  to the UE with probability $P_{b-1,\D}$ which is the target detection probability in block~$b-1$. With probability $1-P_{b-1, \D}$ no URLLC message is generated. If present, the message $m_{b,\U}$ is uniformly drawn from the set $\mathcal M_{\U} : = \{1, \ldots, M_{\U}\}$. 

In each block~$b$, the BS thus creates its dual-function signal $\vectsf X_b \in \mathbb C^{\ell \times t}$ as 
\begin{IEEEeqnarray}{rCl}
\vectsf X_b = \begin{cases} f_b^{(\ell)}(m_e, m_{b,\U}), & \text{with probability}\;  P_{b-1,\D} \\
f_e^{(\ell)}(m_e), & \text{with probability} \; 1- P_{b-1,\D}, \end{cases}
\end{IEEEeqnarray}
where $f_e^{(\ell)}$ and $f_b^{(\ell)}$ are encoding functions on appropriate domains. We assume that $P_{0,\D} = 0$, i.e., no URLLC is transmitted over the first block. 
Denote  $\vectsf X : = [\vectsf X_1^T, \ldots, \vectsf X_{\eta}^T]^T$. The input matrix $\vectsf X$ is admissible if it belongs to the following set:
\begin{equation}
\mathcal P_X (\P) := \left \{ \vectsf X \in \mathbb C^{n \times t} : \text{Tr} \left(\vectsf X \vectsf X^H\right) \le n \P \right \}, \label{eq:1}
\end{equation}
that implies a power constraint on the input matrix $\vectsf X$ by upper bounding the trace of $\vectsf X \vectsf X^H$ with $n \P$.
Define
\begin{IEEEeqnarray}{rCl}\label{eq:barr}
\Barr : = \{ b \in [\eta]: \text{block} \; b \; \text{is with URLLC}\},
\end{IEEEeqnarray}
as the set of blocks in which an additional URLLC message is generated.

\subsection{Communication Received Signal at the UE}
At the end of each block~$b \in [\eta]$, the UE receives the signal $\vectsf Y_{b,c} = [\vect Y_{b,c,1} \ldots \vect Y_{b,c,r} ] \in \mathbb C^{\ell \times r}$ from the BS. Assume a MIMO memoryless Gaussian quasi-static fading channel. The channel input-output relation in each block~$b \in [\eta]$ is given by
\begin{equation} \label{eq:channel}
\vectsf Y_{b,c} =\vectsf X_{b}  \vectsf H_{b,c} \ + \vectsf N_{b,c},
\end{equation}
where   $\vectsf Y_{b,c} \in \mathbb C^{\ell\times r}$ is the communication channel output, $\vectsf H_{b,c} \in \mathbb C^{t \times r}$ is the communication channel matrix, and $ \vectsf N_{b,c} \in \mathbb C^{\ell \times r}$ is additive noise at the UE whose entries are i.i.d. $\mathcal {N} (0,1)$ and is independent of $\vectsf H_{b,c}$. 
Assume that the channel state information is known at both the BS and the UE. 
After each block~$b \in \Barr$, the UE decodes the transmitted URLLC message $m_{b,\U}$. Thus it produces
\begin{equation}
\hat m_{b, \U} = g_{b,u}^{(\ell)} (\vectsf Y_{b,c}, \vectsf H_{b,c}),
\end{equation}
for some decoding function $g_{b,u}^{(\ell)}$ on appropriate domain. 
Otherwise, the UE produces $\hat m_{b, \U}  = 0$ if $b \notin \Barr$. For each message $m_{b,\U}$, the average error probability is defined as
\begin{IEEEeqnarray}{rCl}
\epsilon_{b,\U}&: =& P_{b-1,\D} \Pr [\hat m_{b, \U} \neq m_{b, \U} | b \in \Barr]\notag \\
&& + (1-P_{b-1,\D}) \Pr [\hat m_{b, \U} \neq 0 | b \notin \Barr].
\end{IEEEeqnarray}

After the entire $\eta$ blocks, the UE decodes its desired eMBB message as:
\begin{equation}
\hat m_{\e} = g_{\e}^{(n)} (\vectsf Y_{1,c}, \ldots, \vectsf Y_{\eta,c}, \vectsf H_{1,c}, \ldots, \vectsf H_{\eta,c}).
\end{equation}
where $g_{\e}^{(n)} $ is a decoding function on appropriate domain. 
 The average error probability of the eMBB message $m_e$ is defined as 
 \begin{equation}
 \epsilon_{\e}^{(n)} : = \Pr [ \hat m_{\e} \neq m_{\e}].
\end{equation}

The objective is to optimize the dual-function transmit waveform $\vectsf X$ to  maximize the eMBB transmission rate, while ensuring that the URLLC decoding error probability $\epsilon_{b, \U}$ remains below a threshold and the target detection probability $P_{b,\D}$ exceeds a threshold across all blocks. 
In the following  Proposition~\ref{prop1}, we formulate the corresponding optimization problem as  the rate-reliability-detection trade-off. 


\begin{proposition}\label{prop1}
Given $n, \eta$ and $\P$, let $\R_e := \frac{\log M_{\e}}{n}$ be the eMBB transmission rate.  The \emph{rate-reliability-detection trade-off}  is
\begin{subequations}
\begin{IEEEeqnarray}{rCl}
\max_{f_{\vectsf X}(\vectsf x) \in \mathcal P_X(\P) } \quad && \R_{\e}\\
\text{subject to}\; \quad && \epsilon_{\e}^{(n)} \le \epsilon_{\e},\\
&& \epsilon_{b, \U} \le \epsilon_{\U}, \quad \forall b \in [\eta], \\
&& P_{b, \D}  \ge \P_{\D}, \quad \forall b \in [\eta],
\end{IEEEeqnarray}
\end{subequations}
where $\mathcal P_X(\P)$ is defined in \eqref{eq:1}. 
\end{proposition}

\section{Coding Scheme} \label{sec:coding}
 In this section, we propose a coding scheme to design the dual-function sensing and communication signal $\vectsf X$ in different transmission blocks with the objective of Proposition~\ref{prop1}. Due to URLLC requirements, we perform our analysis in the finite blocklength (FBL) regime \cite{Nikbakht2024, Lv2025, Shen2023}.   In terms of the choice of input distribution, to improve the FBL analysis of  power-constrained channels, our analysis relies on the use of power-shell codebooks. A power-shell codebook of length $\ell$ consists of codewords that are uniformly distributed on the centered $(\ell-1)$-dimensional sphere with  radius $\sqrt{\ell \P}$ where $\P$ is the power constraint. According to Shannon’s observation, the optimal decay of the probability of error near capacity of the point-to-point Gaussian channel is achieved by codewords on the power-shell \cite{Shannon1959}.

 Codewords are generated such that the total transmit power satisfies the power constraint \eqref{eq:1} over the $\eta$ blocks. Therefore, in each block~$b \notin \Barr$, we choose a power sharing parameter $\bsf \in [0,1]$ to allocate $(1-\bsf)\P$  to the sensing task and the remaining $\bsf \P$  to the communication task of transmitting only the eMBB message. In each block~$b \in \Barr$, we choose power sharing parameters $\bc, \bss \in [0,1]$. We allocate $\bc \P$ to the URLLC transmission,  $\bss (1-\bc)\P$ to the  eMBB transmission and the remaining $(1-\bss)(1-\bc)\P$ to the sensing task.

Denote a centered $\ell$-dimensional  sphere of radius $r$ by $\mathsf S_{\ell}( r)$. For each block $b \in [\eta]$ and for each transmit antenna $j \in [t]$, we generate the following codewords that are uniformly distributed on the power-shell.
\begin{itemize}
\item For each $v \in [ M_v]$ and each  realization  $m \in [ M_{\U}]$, generate  codewords $\vect V_{b,j}(m,v)$    by picking them uniformly over $\mathsf S_{\ell}\left(\sqrt{\ell (\bc + \alc^2(1-\bc)) \P}\right)$ with $\alc \in [0,1]$. 
\item For each $s \in [M_{\s}]$ and each realization $ m' \in [M_{\e}]$,  randomly draw two codewords: a codeword $\vect S_{b,j}^{(1)}(m',s)$ uniformly distributed  on $\mathsf S_{\ell}\left (\sqrt{\ell (\bsf + \alsf^2(1-\bsf)) \P}\right)$ with $\alsf \in [0,1]$; and a codeword $\vect S_{b,j}^{(2)}(m',s)$ uniformly distributed  on  $\mathsf S_{\ell} \left (\sqrt{\ell (1-\bc)(\bss + \alss^2(1-\bss)) \P}\right)$, with $ \alss \in [0,1]$.
\item For each $s \in [M_{\s}]$,  randomly draw two codewords: a codeword $\vect X_{b,j}^{(\s,1)}(s)$ uniformly distributed  on $\mathsf S_{\ell} \left (\sqrt{\ell (1-\bsf) \P}\right)$; and a codeword $\vect X_{b,j}^{(\s,2)}(s)$ uniformly distributed  on $\mathsf S_{\ell} \left (\sqrt{\ell(1-\bc)(1-\bss) \P}\right)$. 
\end{itemize}
All codewords are chosen independently of each other. 
We assume that in each block~$b$, the SR has a full knowledge about the auxiliary codewords $\xbjsf$ and $\xbjs$ for all $j \in [q]$. We thus denote this codewords as sensing signals. However, the SR has no knowledge about the URLLC codeword $\vbj$ and the eMBB codewords $\sbjf$ and $\sbjs$. We denote these codewords as communication signals.  In each block~$b \in [\eta]$ and for each transmit antenna $j\in [q]$,   the dual-function signal $\vect X_{b,j}$ is created by superposing the sensing and communication signals using the DPC technique. Specifically, in each block~$b \notin \Barr$,  we employ a DPC technique with parameter $\alsf \in [0,1]$ to precancel the interference of the sensing signal from the eMBB transmission. In each block~$b \in \Barr$ where the BS has both eMBB and URLLC messages to send, we first precancel the interference of  the sensing signal from the eMBB transmission using a DPC with parameter $\alss \in [0,1]$, and then precancel the interference of both eMBB and sensing signals from the URLLC transmission using a DPC  with parameter $\alc \in [0,1]$. 
In the following sections, we explain encoding, decoding and target detection processes in details.  

\subsection{Encoding}\label{sec:enc1}
\subsubsection{Encoding at each block~$b \notin \Barr$} 
In each block~$b \notin \Barr$, for each $j \in [t]$, the BS first picks its sensing signal $\vect X_{b,j}^{(s,1)}(s)$ and then uses  DPC to encode its eMBB message $m_{\e}$  while precanceling the interference of its own sensing signal. Specifically, it chooses an index $s$ such that the 
\begin{equation} \label{eq:xe1}
\vect X_{b,j}^{(\e,1)} :  = \vect S_{b,j}^{(1)} (m_{\e}, s )- \alsf  \vect X_{b,j}^{(\s,1)}
\end{equation}
 lies in the set $\mathcal D(\ell \bsf \P, \zeta_{s,1})$ for a given $\zeta_{\s,1}> 0$ where 
\begin{IEEEeqnarray}{rCl}\label{eq:di}
\mathcal D(a, \zeta) := \left \{ \vect x: a -\zeta \le \left\| \vect x\right\|^2 \le a \right \}. \IEEEeqnarraynumspace
\end{IEEEeqnarray}
For simplicity, we assume that at least one such a codeword exists. If multiple  such codewords  exist, the index $s$ is chosen at random from this set, and the BS sends $\vectsf X_b = [\vect X_{b,1}, \ldots, \vect X_{b, t}]$ with
\begin{equation}
\vect{X}_{b,j}= \vect X_{b,j}^{(\e,1)} + \vect X_{b,j}^{(\s,1)}, \quad j \in [t].
\end{equation}
\subsubsection{Encoding at each block~$b \in \Barr$} 
 In each block~$b \in  \Barr$, the BS has both eMBB and URLLC messages to send. For each $j \in [t]$,  it first picks its sensing signal $\vect X_{b,j}^{(\s,2)}(s)$. It then uses DPC to encode its eMBB message $m_{\e}$ while precanceling the interference of its own sensing signal. More specifically it chooses an index $s$ such that  
\begin{equation} \label{eq:xe1}
\vect X_{b,j}^{(\e,2)} :  = \vect S_{b,j}^{(2)} (m_{\e}, s )- \alss  \vect X_{b,j}^{(\s,2)}
\end{equation}
 lies in the set $\mathcal D \left (\ell \bss(1-\bc) \P, \zeta_{\s,2}\right)$
for a given $\zeta_{\s,2}> 0$. 
  Then it employs DPC to encode $m_{b, \U}$  while precanceling the interference of its own sensing and eMBB signals  $\vect X_{b,j}^{(\e,2)} + \vect X_{b,j}^{(\s,2)}$. 
Specifically, it chooses an index $v$ such that the 
sequence 
\begin{equation} \label{eq:x21}
\vect X_{b,j}^{(\U)} :  = \vect V_{b,j} (m_{b, \U}, v )- \alc  \left (\vect X_{b,j}^{(\e,2)} + \vect X_{b,j}^{(\s,2)}\right)
\end{equation}
 lies in the set $\mathcal D\left (\ell \bc \P, \zeta_u\right )$
for a given $\zeta_u> 0$. 
If multiple  such codewords  exist, indices  $s$ and $v$ are chosen at random from these sets, and the BS sends $\vectsf X_b = [\vect X_{b,1}, \ldots, \vect X_{b, t}]$ with
\begin{equation}
\vect{X}_{b,j}= \vect X_{b,j}^{(\U)} + \vect X_{b,j}^{(\e,2)} + \vect X_{b,j}^{(\s,2)}, \quad j \in [t].
\end{equation}




%




\subsection{Decoding} 
  
\subsubsection{Decoding of URLLC Messages} At the end of each block~$b \in [\eta]$, the UE observes the channel outputs $\vectsf Y_{b,c}$:
\begin{IEEEeqnarray}{rCl}
\vectsf Y_{b,c} = \begin{cases} 
( \vectsf X_{b}^{(\e,1)} + \vectsf X_{b}^{(\s,1)}) \vectsf H_{b,c}+ \vectsf N_{b,c} \quad & \text{if}\; b \notin \Barr  \\
(\vectsf X_{b}^{(\U)} + \vectsf X_{b}^{(\e,2)} + \vectsf X_{b}^{(\s,2)})\vectsf H_{b,c}+ \vectsf N_{b,c} \quad & \text{o.w.} 
 \end{cases}\notag  \\
\end{IEEEeqnarray}
Define the information density metric between $\vectsf y_{b,c}$ and $\vectsf v_b: = [\vect v_{b,1}, \ldots, \vect v_{b,q}]$ by:
\begin{equation} \label{eq:ibu}
i^{(\U)}_b  (\vectsf v_b; \vectsf y_{b,c} ) := \log \frac{f_{\vectsf Y_{b,c}| \vectsf V_{b}} (\vectsf y_{b,c}| \vectsf v_{b})}{f_{\vectsf Y_{b,c}}(\vectsf y_{b,c})}. 
\end{equation}

The UE  then chooses  the pair
\begin{equation}
(m^*,v^*) =\text{arg} \max_{ m, v}  i^{(\U)}_b  (\vectsf v_b; \vectsf y_{b,c} ).
\end{equation}
Given a threshold $\delta_{\U}$, if for this pair 
\begin{equation}
 i^{(\U)}_b  (\vectsf v_b; \vectsf y_{b,c} ) > \delta_{\U}
 \end{equation}
 the UE chooses $(\hat m_{b,\U},\hat v)= (m^*,v^*)$. Otherwise the UE declares that no URLLC message has been received and indicates it by setting $\hat m_{b,\U}=0$. 
Define
\begin{IEEEeqnarray}{rCl} 
\mathcal B_{\text{detect}} &:=& \{b \in [\eta]: \hat m_{b,\U} \neq 0\}, \label{eq:bdet}\\
\mathcal B_{\text{decode}} &: =& \{b \in \mathcal B_{\text{detect}}: \hat m_{b,\U} = m_{b,\U}\},\label{eq:bdec}
\end{IEEEeqnarray}
where $\Bdet$ denotes the set of blocks in which a URLLC message is detected and $\Bdec$ denotes the set of blocks in which a URLLC message is decoded correctly. 

\subsubsection{Decoding the eMBB Message under the TIN Approach} \label{sec:eMBBTIN}
The UE decodes its eMBB message based on the output of the entire $\eta$ blocks. Let $\vectsf Y_c : = [\vectsf Y_{1,c}, \ldots, \vectsf Y_{\eta, c}]$. Given that the URLLC messages interfere on the eMBB transmissions, under this approach, the UE treats URLLC transmissions as noise. Therefore, the decoding of the eMBB message depends on the detection of URLLC messages sent over the $\eta$ blocks. Let $B_{\text{dt}}$ be the realization of the set $ \mathcal B_{\text{detect}}$ defined in \eqref{eq:bdet}. Also, let $\vectsf s_{\e,1} : =  \{\vectsf s_{b}^{(1)}\}_{b \notin B_{\text{dt}} }$, and $\vectsf s_{\e,2} : =  \{\vectsf s_{b}^{(2)}\}_{b \in  B_{\text{dt}} }$. Given $B_{\text{dt}}$, the UE decodes its eMBB message based on the outputs of the entire $n$ channel uses by  looking for an index pair $(m',s')$ such that its corresponding codewords $\left \{ \vectsf s_{\e,2} (m',s'),  \vectsf s_{\e,1} (m',s') \right \}$ maximize 
\begin{IEEEeqnarray}{rCl}
\lefteqn{i^{(\e)}_{\text{TIN}} \left ( \vectsf s_{\e,2} ,  \vectsf s_{\e,1} ;  \vectsf y_c| \mathcal B_{\text{detect}} = B_{\text{dt}} \right):=}\notag \\ &&
 \log \hspace{-0.15cm}\prod_{b\notin  B_{\text{dt}}  }\hspace{0cm}  \frac{f_{\vectsf Y_{b,c}| \vectsf S_{b}^{(1)}} (\vectsf y_{b,c}| \vectsf s_{b}^{(1)})}{f_{\vectsf Y_{b,c}}(\vectsf y_{b,c})} +  \log \hspace{-0.15cm}\prod_{b\in  B_{\text{dt}} } \hspace{0cm}  \frac{f_{\vectsf Y_{b,c} | \vectsf S_{b}^{(2)}} (\vectsf y_{b,c}| \vectsf s_{b}^{(2)})}{f_{\vectsf Y_{b,c}}(\vectsf y_{b,c})}   \IEEEeqnarraynumspace
\end{IEEEeqnarray}
among all codewords
$\left \{ \vectsf s_{\e,2} (m',s'),  \vectsf s_{\e,1} (m',s') \right \}$.

\subsubsection{Decoding the eMBB Message under the SIC Approach} \label{sec:eMBBTIN}
Under this approach, before decoding the eMBB message, the UE first mitigates the interference of correctly decoded URLLC messages over the $\eta$ blocks. Therefore, the decoding of the eMBB message depends on both the detection of URLLC messages sent over the $\eta$ blocks and the decoding of such messages. Let $\Bdc$ be the realization of the set $ \Bdec$ defined in \eqref{eq:bdet}. Also, let $\vectsf s_{\e,1} : =  \{\vectsf s_{b}^{(1)}\}_{b \notin B_{\text{dt}} }$, and $\vectsf s_{\e,2} : =  \{\vectsf s_{b}^{(2)}\}_{b \in  B_{\text{dt}} }$. Given $\Bdt$ and $\Bdc$, the UE decodes its eMBB message based on the outputs of the entire $n$ channel uses by  looking for an index pair $(m',s')$ such that its corresponding codewords $\left \{ \vectsf s_{\e,2} (m',s'),  \vectsf s_{\e,1} (m',s') \right \}$ maximize 
\begin{IEEEeqnarray}{rCl}
\lefteqn{i^{(\e)}_{\text{SIC}} \left ( \vectsf s_{\e,2} ,  \vectsf s_{\e,1} ;  \vectsf y_c| \Bdet = \Bdt, \Bdec = \Bdc \right):=}\notag \\ &&
 \log \hspace{-0.15cm}\prod_{b\notin  \Bdt  }\hspace{0cm}  \frac{f_{\vectsf Y_{b,c}| \vectsf S_{b}^{(1)}} (\vectsf y_{b,c}| \vectsf s_{b}^{(1)})}{f_{\vectsf Y_{b,c}}(\vectsf y_{b,c})} +  \log \hspace{-0.15cm}\prod_{b\in  \Bdt \backslash \Bdc } \hspace{0cm}  \frac{f_{\vectsf Y_{b,c} | \vectsf S_{b}^{(2)}} (\vectsf y_{b,c}| \vectsf s_{b}^{(2)})}{f_{\vectsf Y_{b,c}}(\vectsf y_{b,c})} \notag \\
 && +  \log \hspace{-0.15cm}\prod_{b\in \Bdc } \hspace{0cm}  \frac{f_{\vectsf Y_{b,c} | \vectsf S_{b}^{(2)}, \vectsf V_b} (\vectsf y_{b,c}| \vectsf s_{b}^{(2)}, \vectsf v_b)}{f_{\vectsf Y_{b,c}| \vectsf V_b}(\vectsf y_{b,c}|\vectsf v_b)} \IEEEeqnarraynumspace
\end{IEEEeqnarray}
among all codewords
$\left \{ \vectsf s_{\e,2} (m',s'),  \vectsf s_{\e,1} (m',s') \right \}$.

\subsection{Target Detection}
Following the proposed encoding and decoding schemes, the received signal at the SR in the alternative hypothesis in \eqref{eq:H1} can be reformulated as 
\begin{IEEEeqnarray}{rCl}
 \vectsf Y_{b,s} = \begin{cases} 
   (\vectsf X_{b}^{(\e,1)}+ \vectsf X_{b}^{(\s,1)} )  \vectsf H_{b,s}  + \vectsf N_{b,s}, & b \notin \Barr \\
  (\vectsf X_b^{(\U)} + \vectsf X_{b}^{(\e,2)}+ \vectsf X_{b}^{(\s,2)} )  \vectsf H_{b,s}  + \vectsf N_{b,s}, &\text{o.w.}
\end{cases} \IEEEeqnarraynumspace
\end{IEEEeqnarray}
In each block, the SR has the knowledge of the sensing signal  but  treats the interference of the communication signal as noise. 

\section{Main Results} \label{sec:main}
Our main results are presented in Theorem~\ref{th1} and Theorem~\ref{th2} where we formulate the optimization problem of Proposition~\ref{prop1} under  the TIN and SIC approaches, respectively. Based on our proposed coding scheme in Section~\ref{sec:coding}, the design of the dual-function transmit waveform $\vectsf X$ reduces to selecting the coding parameters $ \vect \alpha:=\{\alc, \alsf, \alss\}$ and $\vect \beta: = \{\bu, \bsf, \bss\}$. Hence, the objective is to maximize the eMBB transmission rate over different value of $\vect \beta$ and $\vect \alpha$ while ensuring that the URLLC decoding error probability $\epsilon_{b, \U}$ remains below a threshold and the target detection probability $P_{b,\D}$ exceeds a threshold across all blocks. To this end, in this section, we first evaluate the URLLC decoding error probability in Lemma~\ref{lemma2}, the eMBB transmission rate under the TIN and SIC approaches in Lemma~\ref{lemma3} and Lemma~\ref{lemma4}, and the target detection probability in Lemma~\ref{lemma1}.  
By combining these lemmas with Proposition~\ref{prop1},  we then prove Theorem~\ref{th1} and Theorem~\ref{th2}. 

\subsection{URLLC Decoding Error Probability Analysis}
When channel state is known at both the BS and the UE, then by performing a singular value decomposition,  the MIMO channel in \eqref{eq:channel} is transferred into the following set of $q:=\min \{t,r\}$ parallel channels 
\begin{IEEEeqnarray}{rCl}
\vect Y_{b,c,j} = \vect X_{b,j} \sqrt{\lambda_{b,j}} + \vect N_{b,c,j},
\end{IEEEeqnarray}
for each $j \in [q]$ and each $b \in [\eta]$. Here, $\lambda_{b,1} \ge \lambda_{b,2} \ge \ldots \ge \lambda_{b,q}$ are the largest $q$ eigenvalues of $\vectsf H_{b,c} \vectsf H_{b,c}^H$ and $\vect N_{b,c,j} \sim \mathcal{N}(\vect 0, \vectsf I_{\ell \times \ell})$ are independent noise vectors.  Let $\vect \lambda_b := [\lambda_{b,1}, \ldots, \lambda_{b,q}]$. 
Also, let  $\vectsf X_b : = [\vect X_{b,1}, \ldots, \vect X_{b,q}]$. 
In each block~$b \in \Barr$, we have the following URLLC decoding error events:
\begin{eqnarray}
\mathcal E_{\U,1}&:=& \{b \notin \Bdet | b \in \Barr \}, \label{eq:eu1}\\
\mathcal E_{\U, 2}&:=& \{b \notin \Bdec | b \in \Bdet,  b \in \mathcal B_{\text{arrival}}\}, \label{eq:eu2}
\end{eqnarray}
which indicate a missed-detection event. Specifically, $\mathcal E_{\U,1}$ happens when the transmitted URLLC message is not detected and $\mathcal E_{\U,2}$ happens when the decoded URLLC message does not match the transmitted one. 
In each block $b \notin \Barr$, we have the following error event:
\begin{eqnarray}
\mathcal E_{\U,3}&:=& \{b \in \Bdet | b \notin \Barr \}, \label{eq:eu3}
\end{eqnarray}
which indicates a false-alarm event. More specifically, $\mathcal E_{\U,3}$ happens when the UE incorrectly declares the detection of a URLLC message in a block with no URLLC. 
The URLLC decoding error probability thus is bounded as 
\begin{eqnarray}
\epsilon_{b,\U} \le P_{b-1, \D} \left (\Pr [\mathcal E_{\U,1}] + \Pr [\mathcal E_{\U,2}] \right)+ (1- P_{b-1, \D} ) \Pr [\mathcal E_{\U,3}], \label{eq:57}
\end{eqnarray}
where $P_{b-1, \D}$ is the target detection probability in the previous block $b-1$ and is calculated in Lemma~\ref{lemma1}. 
In Appendix~\ref{App:B}, we analyze each error event which results in Lemma~\ref{lemma2}.
\begin{lemma} \label{lemma2}
The average URLLC decoding error probability is upper bounded as
\begin{eqnarray}\label{eq:boundu}
\epsilon_{b,\U} \le  P_{b-1, \D} P_{\U,1}   + (1-P_{b-1, \D})P_{\U,2}
\end{eqnarray}
where 
\begin{eqnarray}
P_{\U,1}&: =& \tepf+ (\tepf) ^{M_{\U} M_v}+\teps,\\
P_{\U,2}&: =& 1 - \left (1-\frac{\teps}{M_{\U}M_{v}}\right)^{M_{\U} M_v} ,
\end{eqnarray}
\begin{IEEEeqnarray}{rCl}
\tilde \epsilon_{\U,1} &: =& Q \left (\frac{ \ell \C_{\U}-\log(M_{\U} M_v)  - K_{\U} \log (\ell)}{\sqrt{\ell \mathsf V_{\U}}} \right) + \frac{B}{\sqrt{\ell} }, \IEEEeqnarraynumspace \label{eq:tepf} \\
\tilde \epsilon_{\U,2} &:=& \frac{2}{\ell^{K_{\U} }} \left ( \frac{\log 2}{\sqrt{2\pi \ell}} + \frac{2B } {\sqrt{\ell}}\right),\label{eq:teps}
\end{IEEEeqnarray}
 for some $K_{\U}>0$ and $B >0$ and where $Q(x) = \frac{1}{\sqrt{2\pi}} \int_{x}^{\infty} \exp \left (-\frac{t^2}{2}\right) dt$ is the Q-function and
\begin{IEEEeqnarray}{rCl}\label{eq:CuVu}
\C_{\U} : =  \sum_{j = 1}^q \C(\Omega_{b,j}),\quad 
\mathsf V_{\U} := \sum_{j = 1}^q \mathsf V(\Omega_{b,j}),
\end{IEEEeqnarray}
with $\C(x) = \frac{1}{2} \log (1+x)$, $\mathsf V(x): = \frac{x(2+x)}{2(1+x)^2}$ 
with
\begin{IEEEeqnarray}{rCl}
\Omega_{b,j}&: =& \frac{\sigy - \sigyvj }{\sigyvj}, \label{eq:omegabj} \\
\sigy &:=& 1+\lambda_{b,j} \P, \label{eq:sigy}\\
\sigyvj &:=& 1+\lambda_{b,j}(1-\alc^2)(1-\bc)\P. \label{eq:sigyvj}
\end{IEEEeqnarray}

\end{lemma}
\begin{IEEEproof}
See  Appendix~\ref{App:B}. 
\end{IEEEproof}
\begin{remark}\label{rem2}
Under the assumption that $\Pr[\mathcal E_{\U,1}] \to 0$ and $\Pr[\mathcal E_{\U,3}] \to 0$, i.e., when the probability that all the URLLC messages are detected correctly is almost $1$, then for sufficiently large $\ell$,
\begin{IEEEeqnarray}{rCl}
\epsilon_{b,\U} &\le&  P_{b-1, \D} (\tepf + \teps) \notag \\
&=& P_{b-1, \D} Q \left (\frac{ \ell \C_{\U}-\log(M_{\U} M_v)  - O( \log (\ell))}{\sqrt{\ell \mathsf V_{\U}}} \right),
\end{IEEEeqnarray}
 where $\C_{\U}$ and $\mathsf V_{\U}$ are defined in \eqref{eq:CuVu}. This result is consistent with the FBL analysis of decoding error probability of parallel AWGN channels proposed in \cite[Theorem~78]{YuryThesis}. 
\end{remark}
\begin{remark}\label{rem3}
In the single-input single-output (SISO) case, the bound on the URLLC decoding error probability can be recovered from Lemma~\ref{lemma2} by setting $q = 1$ and replacing $\lambda_{b,j}$ with $\lambda_b$ for all $j \in [q]$. In this case, under the assumptions of Remark~\ref{rem2}, we have 
\begin{IEEEeqnarray}{rCl}
\epsilon_{b,\U} &\le&  P_{b-1, \D} Q \left (\frac{ \ell \C(\Omega_{b,j})-\log(M_{\U} M_v)  -  O( \log (\ell))}{\sqrt{\ell \mathsf V(\Omega_{b,j})}} \right), \IEEEeqnarraynumspace
\end{IEEEeqnarray}
\end{remark}
where $\C(x) = \frac{1}{2} \log (1+x)$, $\mathsf V(x): = \frac{x(2+x)}{2(1+x)^2}$. This result is consistent with the FBL analysis of decoding error probability of point-to-point single AWGN channels proposed by Polyanskiy, Poor and Verdú in \cite[Theorem~54]{Yuri2012}.
\subsection{eMBB Transmission Rate Analysis Under the TIN Approach} 
The eMBB message is decoded at the end of the entire $n$ channel uses. Under this approach, the UE treats URLLC transmissions as noise. Therefore, the decoding of the eMBB message depends on the detection of URLLC messages sent over the $\eta$ blocks. We have the following two eMBB decoding error events:
\begin{eqnarray}
\mathcal E_{\e,1} &: =& \{\Bdet \neq \Barr\}, \label{eq:Ee1} \\
\mathcal E_{\e,2} &: =& \{\hat m_{\e} \neq m_e | \Bdet = \Barr\}, \label{eq:Ee2}
\end{eqnarray}
where $\mathcal E_{\e,1} $ happens when the UE does not successfully detect all the transmitted URLLC messages, and $\mathcal E_{\e,2}$ happens when the UE successfully detects all the transmitted URLLC messages but the decoded eMBB messages does not match the transmitted one. 

 Recall the definition of $B_{\text{dt}}$ as  the realization of the set $ \mathcal B_{\text{detect}}$. The average eMBB decoding error probability is bounded by
\begin{IEEEeqnarray}{rCl} \label{eq:51}
\epsilon_{\e, \text{TIN}}^{(n)} &\le&
\sum_{B_{\text{dt}}} \Pr [\Bdet = B_{\text{dt}}] \notag \\
&& \Big ( \Pr [\mathcal E_{\e,1}| \Bdet = B_{\text{dt}}] + \Pr [\mathcal E_{\e,2} | \Bdet = B_{\text{dt}}] \Big).
\end{IEEEeqnarray}
In Appendix~\ref{App:C}, by analyzing the occurrence probability of each error event, we calculate the right-hand side (RHS) of \eqref{eq:51}.  We then use  the Berry-Esseen central limit theorem (CLT) for functions \cite[Proposition~1]{MolavianJaziArXiv} to upper bound the eMBB transmission rate which results in the following lemma.

\begin{lemma}\label{lemma3}
Under the TIN approach, the eMBB transmission rate $\R_{\e}:= \frac{\log M_{\e}}{n}$ is upper bounded as
\begin{eqnarray}
\R_{\e} \le  {\C}_{\e} - \sqrt{ \frac{ \mathsf V_{\e}}{n}}Q^{-1}(\epsilon_{\e}-\Delta_{\e}) - K_{\e} \frac{\log (n)}{n} - \frac{\log(M_{\s})}{n}
\end{eqnarray}
for some $K_{\e} > 0$ and where 
\begin{IEEEeqnarray}{rCl}
\C_{\e} &:=&\sum_{\Bdt} \P_{\text{dt}}^{|\Bdt|} (1-\P_{\text{dt}})^{\eta-|\Bdt|} \tilde {\C}_{\e}, \label{eq:Ce}\\
\mathsf V_{\e} &:=& \sum_{\Bdt} \P_{\text{dt}}^{|\Bdt|} (1-\P_{\text{dt}})^{\eta-|\Bdt|} \tilde {\mathsf V}_{\e}, \label{eq:Ve}\\
\tilde{\C}_{\e}&:=&\frac{1}{\eta}\sum_{j = 1}^q \left (\sum_{b \notin \Bdt} \C(\Omega_{b,j}^{(1)}) + \sum_{b \in \Bdt} \C( \Omega_{b,j}^{(2)})\right),\label{eq:tce}\\
\tilde{\mathsf V}_{\e}&:=&\frac{1}{\eta} \sum_{j = 1}^q \left (\sum_{b \notin \Bdt} \mathsf V(\Omega_{b,j}^{(1)}) + \sum_{b \in \Bdt} \mathsf V(\Omega_{b,j}^{(2)})\right),\label{eq:tve}
\end{IEEEeqnarray}
where $\C(x) = \frac{1}{2} \log (1+x)$, $\mathsf V(x): = \frac{x(2+x)}{2(1+x)^2}$,
\begin{IEEEeqnarray}{rCl}
\P_{\text{dt}}&: =& P_{b-1,\D} (1- (\hat \epsilon_{\U,1})^{M_{\U}M_v}) +  (1-P_{b-1,\D}) P_{\U,2}, \label{eq:Pdtn}\\
\Delta_{\e}&:=& 1 + \frac{\tilde B}{\sqrt{n}}\left ( 1 + \frac{4}{n^{K_{\e}}}\right)  +\frac{2\log 2}{n^{K_{\e}}\sqrt{2n\pi}}\notag \\
&&-\sum_{\Bdt} \left (P_{b-1, \D}(1-(\tepf)^{M_{\U}M_v})\right)^{|\Bdt|}\notag \\
&& \hspace{1cm} \times \left ((1-P_{b-1, \D}) (1-P_{\U,2}) \right)^{\eta - |\Bdt|}, \label{eq:Deltae} \IEEEeqnarraynumspace\\
\Omega_{b,j}^{(1)}&: =& \frac{\sigy - \sigysfj}{\sigysfj},\quad 
\Omega_{b,j}^{(2)} := \frac{\sigy - \sigyssj}{\sigyssj}, \label{eq:Omegas}
\end{IEEEeqnarray}
 with $\hat \epsilon_{\U,1}:= \tepf - \frac{2\tilde B}{\sqrt{\ell}}$, and where $\tepf$ and $\teps$ are defined in \eqref{eq:tepf} and \eqref{eq:teps}, respectively, and
\begin{IEEEeqnarray}{rCl} 
\sigysfj &:=&1 + \lambda_{b,j} (1-\alsf^2)(1-\bsf) \P, \label{eq:sigysfj} \\
\sigyssj &:=& 1 + \lambda_{b,j} \P\Big(1-(1-\alc)^2(1-\bc)\notag \\
&&\hspace{1.5cm} (1-(1-\alss^2)(1-\bss))\Big). \label{eq:sigyssj} \IEEEeqnarraynumspace
\end{IEEEeqnarray}
  \end{lemma}
\begin{IEEEproof}
See  Appendix~\ref{App:C}.
\end{IEEEproof}
\begin{remark}\label{rem4}
The parameter  $P_{\text{dt}}$ can be interpreted as the probability of correct detection of a URLLC message in a given block~$b$. In the case where $\lambda_{b,j} = \lambda$ for all $b \in [\eta]$ and all $j \in [q]$, the capacity and the channel dispersion terms can be simplified  as
\begin{IEEEeqnarray}{rCl}
{\C}_{\e} &=& q \left ( (1-\P_{\text{dt}}) \C(\Omega^{(1)}) + \P_{\text{dt}} \C(\Omega^{(2)})\right), \\
\mathsf V_{\e} &=& q \left ( (1-\P_{\text{dt}}) \mathsf V(\Omega^{(1)}) + \P_{\text{dt}} \mathsf V(\Omega^{(2)})\right), 
\end{IEEEeqnarray} 
where $\Omega^{(1)}$ and $\Omega^{(2)}$ are obtained from \eqref{eq:Omegas} by replacing $\lambda_{b,j}$ with $\lambda$. 
As a result, under the TIN approach, the eMBB transmission channel is interpreted as $q$ parallel AWGN channels where in each channel, with probability $P_{\text{dt}}$ a URLLC message is detected and with  probability $1- P_{\text{dt}}$, no URLLC message is detected. 
\end{remark}

\subsection{eMBB Transmission Rate Analysis Under the SIC Approach} 
Under this approach, before decoding the eMBB message, the UE first mitigates the interference of correctly decoded URLLC messages over the $\eta$ blocks. Therefore, the decoding of the eMBB message depends on both the detection of URLLC messages sent over $\eta$ blocks and the correct decoding of such messages. 
 Recall the definition of $\Bdt$ as a realization of the set $ \Bdet$ and the definition of $\Bdc$ as a realization of the set $\Bdec$. To decode the eMBB message, we again have the two error events defined in \eqref{eq:Ee1} and \eqref{eq:Ee2}. The average eMBB decoding error probability is bounded as
\begin{IEEEeqnarray}{rCl} \label{eq:61}
\epsilon_{\e, \text{SIC}}^{(n)} &\le&
\sum_{B_{\text{dt}}} \Pr [\Bdet = B_{\text{dt}}] \Big ( \Pr [\mathcal E_{\e,1}| \Bdet = B_{\text{dt}}] \notag \\
&&+\sum_{\Bdc} \Pr[\Bdec = \Bdc|\Bdet = \Barr, \Bdet = \Bdt]  \notag \\
&& \hspace{0.5cm}\Pr [\mathcal E_{\e,2} | \Bdet = \Bdt , \Bdec = \Bdc] \Big). \label{eq:71}
\end{IEEEeqnarray}
Note that, compared to the TIN approach, here the decoding of the eMBB message not only depends on the detected URLLC messages set $\Bdt$ but also depends on the correctly decoded URLLC messages set $\Bdc \subset \Bdt$. 
Under this approach, we have the following lemma on the eMBB transmission rate. 
\begin{lemma}\label{lemma4}
Under the SIC approach, the eMBB transmission rate $\R_{\e}:= \frac{\log M_{\e}}{n}$ is upper bounded by
\begin{equation}
\R_{\e} \le  {\C}_{\e,2} - \sqrt{ \frac{ \mathsf V_{\e,2}}{n}}Q^{-1}(\epsilon_{\e}-\Delta_{\e}) - K_{\e} \frac{\log (n)}{n} - \frac{\log(M_{\s})}{n}
\end{equation}
for some $K_{\e} > 0$ and where 
\begin{IEEEeqnarray}{rCl}
\C_{\e,2} &:=& \sum_{\Bdt}\sum_{\Bdc} \mathcal A (\P_{\text{dt}}, \P_{\text{dc}})\tilde {\C}_{\e,2}, \label{eq:Ce2}\\ 
\mathsf V_{\e,2} &:=& \sum_{\Bdt}\sum_{\Bdc} \mathcal A (\P_{\text{dt}}, \P_{\text{dc}})\tilde {\mathsf V}_{\e,2},\label{eq:Ve2} \\
\tilde{\C}_{\e,2}&:=&\frac{1}{\eta}\sum_{j = 1}^q \hspace{-0.1cm} \left (\hspace{-0.05cm} \sum_{b \notin \Bdt} \hspace{-0.15cm}\C(\Omega_{b,j}^{(1)}) +\hspace{-0.3cm} \sum_{b \in \Bdt \backslash \Bdc} \hspace{-0.3cm}\C( \Omega_{b,j}^{(2)}) + \hspace{-0.15cm}  \sum_{b \in \Bdc} \hspace{-0.15cm}  \C( \Omega_{b,j}^{(3)})\right),\label{eq:tce2} \notag \\ \\
\tilde{\mathsf V}_{\e,2}&:=&\frac{1}{\eta} \sum_{j = 1}^q \hspace{-0.1cm} \left (\hspace{-0.05cm} \sum_{b \notin \Bdt} \hspace{-0.15cm}\mathsf V(\Omega_{b,j}^{(1)}) +\hspace{-0.3cm} \sum_{b \in \Bdt \backslash \Bdc} \hspace{-0.3cm}\mathsf V( \Omega_{b,j}^{(2)}) + \hspace{-0.15cm}  \sum_{b \in \Bdc} \hspace{-0.15cm}  \mathsf V( \Omega_{b,j}^{(3)})\right),\notag \\ \label{eq:tve2}
\end{IEEEeqnarray}
where $\C(x) = \frac{1}{2} \log (1+x)$, $\mathsf V(x): = \frac{x(2+x)}{2(1+x)^2}$,
\begin{equation}
\mathcal A (a, b)  := a^{|\Bdt|} (1-a)^{\eta-|\Bdt|} b^{|\Bdc|} (1-b)^{|\Bdt|- |\Bdc|},
\end{equation}
and 
\begin{IEEEeqnarray}{rCl}
\P_{\text{dc}}&: =& 1- \tepf - \teps,\\
\Omega_{b,j}^{(3)}&: =& \frac{\sigyvj - \sigysvj}{\sigysvj} \label{eq:Omega3}
\end{IEEEeqnarray}
with
\begin{IEEEeqnarray}{rCl} 
\lefteqn{\sigysvj} \notag \\
 &:=&1 + \lambda_{b,j} (1-\alc)^2(1-\alsf)^2(1-\bc)(1-\bsf) \P, \label{eq:sigysvj}
\end{IEEEeqnarray}
and $\tepf$ and $\teps$ are defined in \eqref{eq:tepf} and \eqref{eq:teps}, respectively. 
\end{lemma}
\begin{IEEEproof}
See  Appendix~\ref{App:D}.
\end{IEEEproof}

\begin{remark}\label{rem5}
In Lemma~\ref{lemma4},  $P_{\text{dc}}$ can be interpreted as the probability of correctly decoding a URLLC message in each block~$b$. In the case where $\lambda_{b,j} = \lambda$ for all $b \in [\eta]$ and all $j \in [q]$, the capacity and the channel dispersion terms can be simplified  as
\begin{IEEEeqnarray}{rCl}
{\C}_{\e,2} &=& q  (1-P_{\text{dt}}) \C(\Omega^{(1)}) \notag \\
&& + q P_{\text{dt}} \left ((1-P_{\text{dc}}) \C(\Omega^{(2)}) + P_{\text{dc}} \C(\Omega^{(3)}) \right), \\
\mathsf V_{\e,2} &=& q  (1-P_{\text{dt}}) \mathsf V(\Omega^{(1)}) \notag \\
&& + q P_{\text{dt}} \left ((1-P_{\text{dc}}) \mathsf V(\Omega^{(2)}) + P_{\text{dc}} \mathsf V(\Omega^{(3)} \right), 
\end{IEEEeqnarray} 
where $\Omega^{(1)}$, $\Omega^{(2)}$ and $\Omega^{(3)}$ are obtained from \eqref{eq:Omegas} and \eqref{eq:Omega3} by replacing $\lambda_{b,j}$ with $\lambda$. 
As a result, under the SIC approach, the eMBB transmission channel is interpreted as $q$ parallel AWGN channels where in each channel, with probability $1- P_{\text{dt}}$, no URLLC message is detected, with probability $P_{\text{dt}}P_{\text{dc}}$ a URLLC message is correctly detected and decoded, and with probability  $P_{\text{dt}}(1-P_{\text{dc}})$ the correctly detected URLLC message is missed decoded.  
\end{remark}

\subsection{Target Detection Probability Analysis}
Given that the sensing channel state is known at the SR, the MIMO channel in \eqref{eq:H1} is transferred into the following set of $q$ parallel channels 
\begin{IEEEeqnarray}{rCl}
\vect Y_{b,s,j} = \vect X_{b,j} \sqrt{\gamma_{b,j}} + \vect N_{b,s,j},
\end{IEEEeqnarray}
for each $j \in [q]$ and each $b \in [\eta]$. 
Here, $\gamma_{b,1} \ge \gamma_{b,2} \ge \ldots \ge \gamma_{b,q}$ are the $q$ largest eigenvalues of $\vectsf H_{b,s}^H \vectsf H_{b,s} $.  The LRT test in each block~$b \in [\eta]$  is defined in \eqref{eq:test}. 
Given that $\vect N_{b,s,j} \sim \mathcal N(0, \vectsf I_{\ell \times \ell})$, thus $f\left(\vect Y_{b,s,j}  | \mathcal H_0\right)$ is a Gaussian distribution. However, due to our power-shell code construction, this is not the case for $f\left(\vect Y_{b,s,j}  | \mathcal H_1\right)$. Hence, we take a change of metric measurement approach by introducing the following new LRT test: 
\begin{equation}\label{eq:test2}
\tilde T_b := \log  \left ( \prod_{j = 1}^{q} \frac{Q\left (\vect Y_{b,s,j}  | \mathcal H_1\right)}{f\left(\vect Y_{b,s,j}  | \mathcal H_0\right)} \right) \underset{{\mathcal H_0}}{\overset{\mathcal H_{1}}{\gtrless}} \delta,
\end{equation}
where $Q\left(\vect Y_{b,s,j}  | \mathcal H_1\right)$ is a Gaussian distribution and $\vect y_{b,s,j}| \mathcal H_1 \sim \mathcal N(\vect \mu_{b,j}, \sigbj \vect I_{\ell})$ with 
\begin{IEEEeqnarray}{rCl}
\vect \mu_{b,j} &:=& \begin{cases}
\sqrt{\gamma_{b,j}}(1-\alsf)\vect x_{b,j}^{(\s,1)}, \quad & b \notin \Barr \\
\sqrt{\gamma_{b,j}}(1-\alc)(1-\alss)\vect x_{b,j}^{(\s,2)}, \quad & \text{o.w.}
\end{cases} \IEEEeqnarraynumspace\\
\sigbj &:=& 1+ \gamma_{b,j}\kappa_2 (1-\kappa_1) \P \label{eq:sigbj}
\end{IEEEeqnarray}
where
\begin{IEEEeqnarray}{rCl}
\kappa_1&:=& \begin{cases} (1-\alsf^2)(1-\bsf), & b \notin \Barr \\
(1-\alss^2)(1-\bss),  & \text{o.w.}
\end{cases}\label{eq:kappa1}\\
\kappa_2&:=& \begin{cases} 1 , & b \notin \Barr \\
(1-\bc)(1-\alc)^2,  & \text{o.w.}
\end{cases}\label{eq:kappa2}
\end{IEEEeqnarray}


The test thus can be written as
\begin{IEEEeqnarray}{rCl}
\tilde T_b = \sum_{j = 1}^q \left (|| \vect Y_{b,s,j}||^2 -\frac{ ||\vect Y_{b,s,j} -  \vect \mu_{b,j} ||^2}{\sigbj}\right)\underset{{\mathcal H_0}}{\overset{\mathcal H_{1}}{\gtrless}} \delta. \IEEEeqnarraynumspace \label{eq:test3}
\end{IEEEeqnarray}
Accordingly, in each block~$b$, the target detection and false alarm probabilities are given by
\begin{IEEEeqnarray}{rCl}
P_{b,\D} &=& \Pr [\tilde T_b > \delta | \mathcal H_1], \quad 
P_{b,\text{FA}}  =  \Pr [\tilde T_b > \delta | \mathcal H_0].
\end{IEEEeqnarray}

We set $\delta$ such that $P_{b,\text{FA}}  = P_{\text{FA}}$ across all $\eta$ blocks.   The target detection probability then is given by the following lemma. 
\begin{lemma} \label{lemma1}
The target detection probability is given by
\begin{IEEEeqnarray}{rCl} \label{eq:pbD}
P_{b, \D} = 1- F_{\tilde{\chi}_2}  \left (F^{-1}_{\tilde{\chi}_1} (1-P_{\text{FA}}) \right),
\end{IEEEeqnarray}
where $P_{\text{FA}}$ is the desired false alarm probability, $\tilde {\chi}_1 (\{ w_{b,j}\}_{j =1}^{q}, \ell, \{ \nu_{b,j}\}_{j = 1}^q)$ and $\tilde {\chi}_2(\{ w_{b,j}\}_{j =1}^{q}, \ell, \{\tilde \nu_{b,j}\}_{j = 1}^q)$ are generalized chi-square distributions with $F_{\tilde{\chi}_1}(\cdot)$ and $F_{\tilde{\chi}_2}(\cdot)$ as their corresponding cumulative distribution functions (CDFs) and
\begin{subequations}\label{eq:27}
\begin{IEEEeqnarray}{rCl}
w_{b,j}&:=&1-\frac{1}{\sigbj},\\
 \nu_{b,j} &:=&\kappa_1 \kappa_2 \ell \P \left (\frac{1-w_{b,j}}{w_{b,j}}\right)^2, \IEEEeqnarraynumspace\\
\tilde \nu_{b,j}&:=& \begin{cases}
\gamma_{b,j} \ell \P \tau_2(\bsf, \alsf)& b \notin \Barr, \\
\gamma_{b,j} \ell \P \tau_3& \text{o.w.}
\end{cases}
\end{IEEEeqnarray}
\end{subequations}
where 
\begin{IEEEeqnarray}{rCl}
\tau_2(b,a)&:=& b + (1-b)\left(a + \frac{\sigbj}{\sigbj-1}(1-a)\right)^2, \\
\tau_3&: =& \bc + (1-\bc)(\alc^2 + (1-\alc)^2 \tau(\bss,\alss)), \IEEEeqnarraynumspace
\end{IEEEeqnarray}
%
and $\sigbj$, $\kappa_1$ and $\kappa_2$ are defined in \eqref{eq:sigbj}, \eqref{eq:kappa1} and \eqref{eq:kappa2}, respectively.  
\end{lemma}
\begin{IEEEproof}
See  Appendix~\ref{App:A}.
\end{IEEEproof}
\begin{remark}\label{rem1}
In the SISO case, the target detection probability can be recovered from Lemma~\ref{lemma1}. In this case, $\tilde{\chi}_1$ and $\tilde{\chi}_2$ are non-central chi-square distributions with degree of $\ell$ and parameters $\nu_b$ and $\tilde \nu_b$, respectively. These parameters can be calculated from \eqref{eq:27} by assuming $\gamma_{b,j} = \gamma_b$ for all $j \in [q]$.
\end{remark} 

\subsection{Rate-Reliability-Detection Trade-off}
By combining Lemmas~\ref{lemma2}, \ref{lemma3} and \ref{lemma1} with Proposition~\ref{prop1}, we have the following theorem on the rate-reliability-detection trade-off under the TIN approach.
\begin{theorem}\label{th1}
Given $n$ and $\P$, the rate-reliability-detection trade-off under the TIN approach is given by 
\begin{subequations}
\begin{IEEEeqnarray}{rCl}
\max_{\vect \beta, \vect \alpha} \;&& \C_{\e} - \sqrt{ \frac{\mathsf V_{\e}}{n}}Q^{-1}(\epsilon_{\e} - \Delta_{\e}) - K_{\e} \frac{\log (n)}{n} - \frac{\log(M_{\s})}{n}\label{eq:embb}\\ \notag \\
\text{s.t.:}\;  && P_{b-1, \D} P_{\U,1}   + (1-P_{b-1, \D})P_{\U,2} \le \epsilon_{\U}, \quad \forall b \in [\eta], \label{eq:urllc}\IEEEeqnarraynumspace\\ \notag \\
&& 1- F_{\tilde{\mathcal X_2}} (F^{-1}_{\tilde{\mathcal X_1}} (1-P_{\text{FA}})) \ge \P_{\D}, \quad \forall b \in [\eta ], \label{eq:pdt}
\end{IEEEeqnarray}
\end{subequations}
where $\vect \beta:=\{\bc, \bsf,\bss\}$ and $\vect \alpha:=\{\alc,\alsf,\alss\}$.  
\end{theorem}

By combining Lemmas~\ref{lemma2}, \ref{lemma4}, and \ref{lemma1} with Proposition~\ref{prop1}, we have the following theorem on the rate-reliability-detection trade-off under the SIC approach.
\begin{theorem}\label{th2}
Given $n$ and $\P$, the rate-reliability-detection trade-off under the SIC approach is given by 
\begin{subequations}
\begin{IEEEeqnarray}{rCl}
\max_{\vect \beta, \vect \alpha} \;&& \C_{\e,2} - \sqrt{ \frac{\mathsf V_{\e,2}}{n}}Q^{-1}(\epsilon_{\e} - \Delta_{\e,2}) - K_{\e,2} \frac{\log (n)}{n} - \frac{\log(M_{\s})}{n} \notag \\ \\
\text{s.t.:}\;  && \text { \eqref{eq:urllc} and \eqref{eq:pdt}}
\end{IEEEeqnarray}
\end{subequations}
where $\vect \beta:=\{\bc, \bsf,\bss\}$ and $\vect \alpha:=\{\alc,\alsf,\alss\}$.  
\end{theorem}
\begin{figure}[t]
\center
				\includegraphics[width=0.37\textwidth]{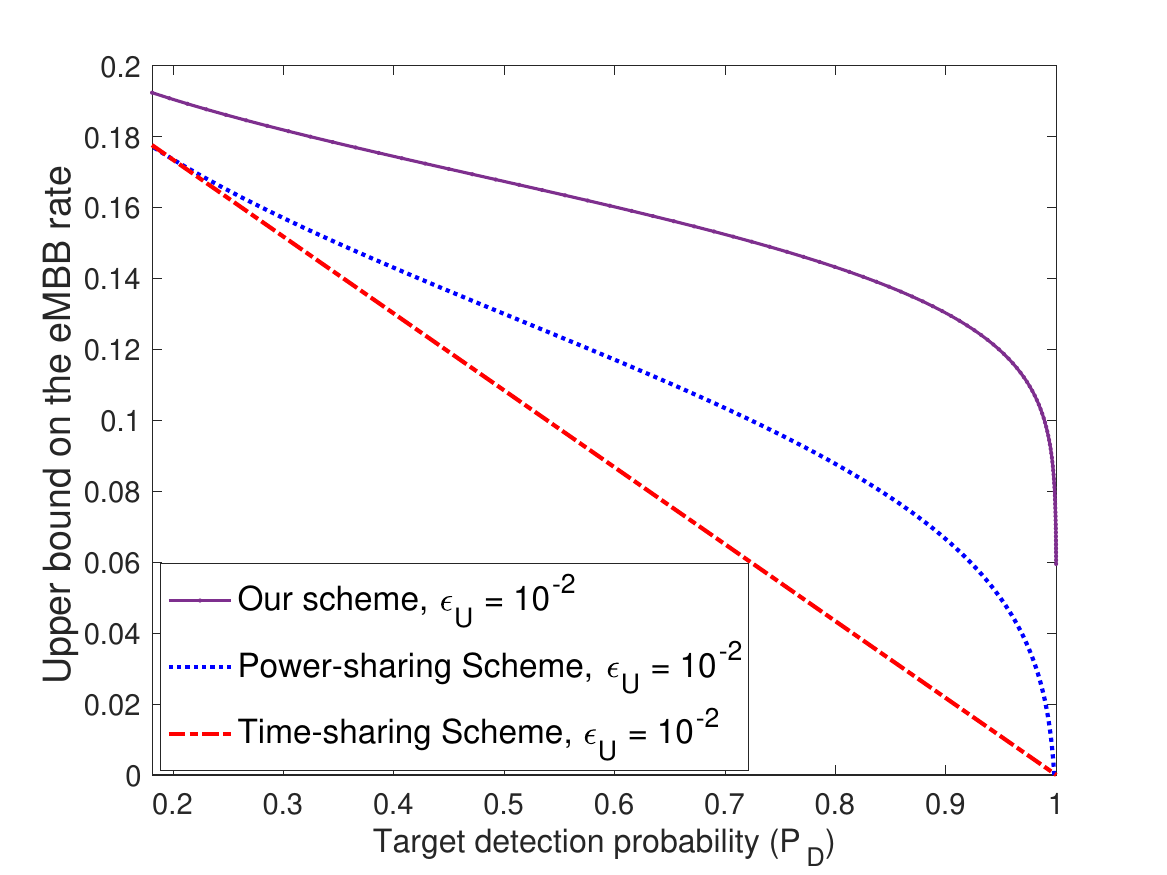}
\caption{Comparing our DPC-based scheme under the TIN approach with power-sharing and time-sharing schemes for $\P= 0.5$, $\ell = 150$, $\eta = 10$, $P_{FA} = 10^{-6}$, $\epsilon_e = 10^{-3}$. }
\label{fig2}
\vspace{-0.4cm}
\end{figure}

\begin{figure}[t]
\center
				\includegraphics[width=0.35\textwidth]{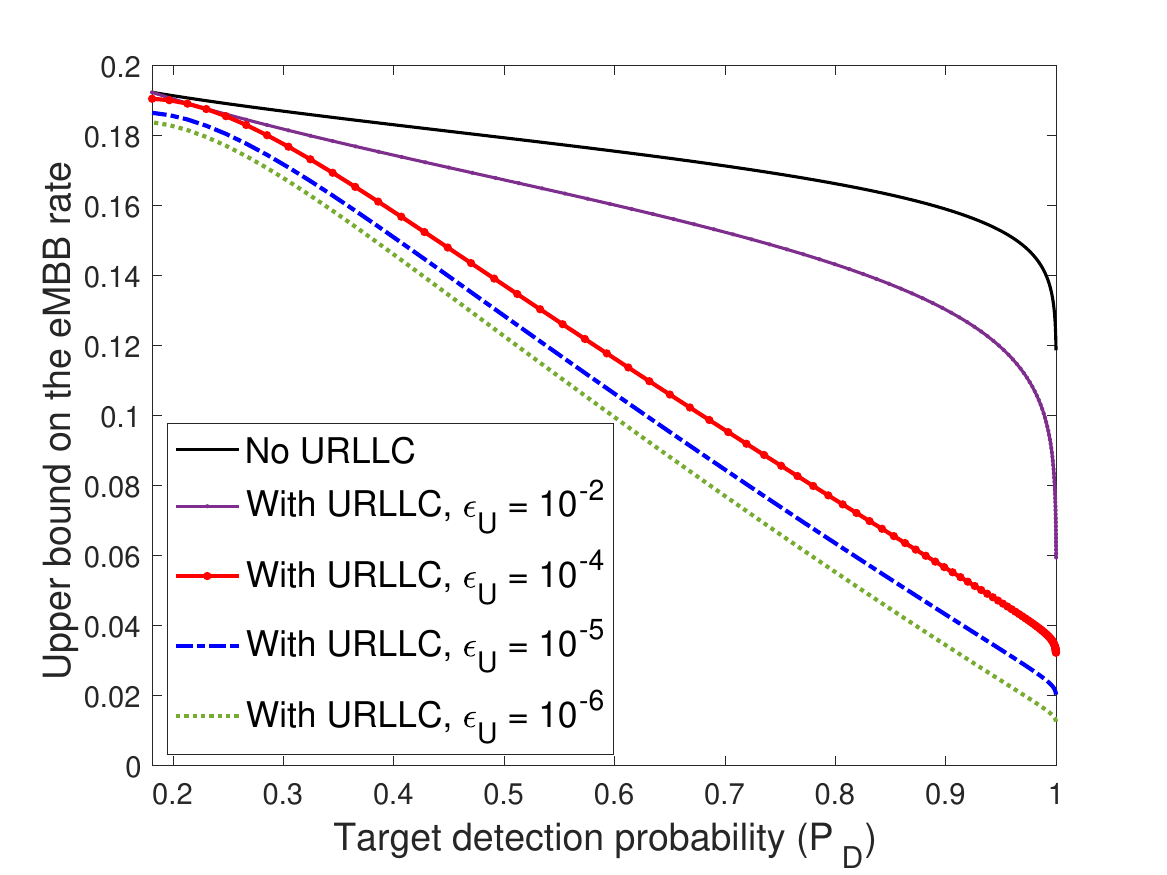}
\caption{Our scheme under the TIN approach with  different values of the URLLC decoding error probability threshold for $\P= 0.5$, $\ell = 150$, $\eta = 10$, $P_{FA} = 10^{-6}$, $\epsilon_e = 10^{-3}$.}
\label{fig3}
\vspace{-0.4cm}
\end{figure}

\begin{figure}[t]
\center
				\includegraphics[width=0.35\textwidth]{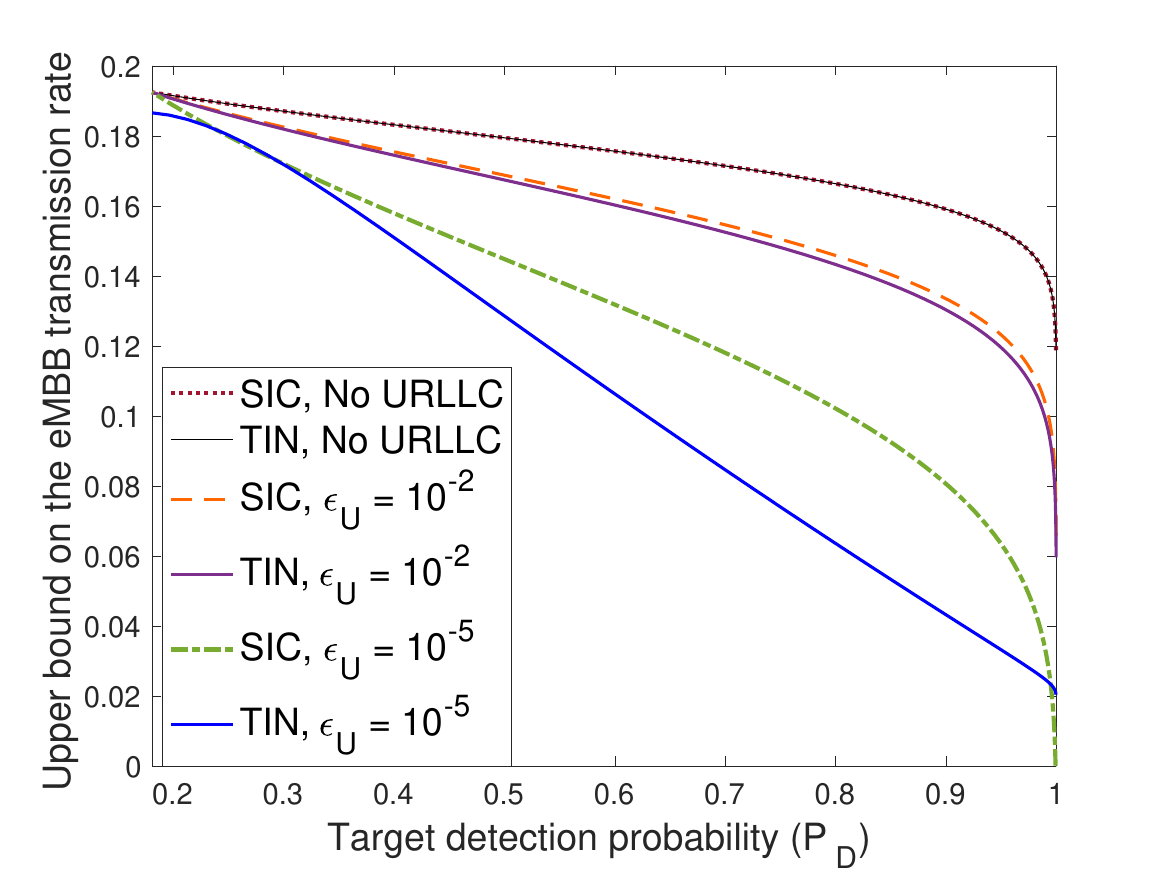}
\caption{Comparing the upper bound on the eMBB transmission rate under the TIN and  SIC approaches for $\epsilon_{\U}$ set at $10^{-2}$ and $10^{-5}$ and  $\P= 0.5$, $\ell = 150$, $\eta = 10$, $P_{FA} = 10^{-6}$, $\epsilon_e = 10^{-3}$.}
\label{fig4}
\vspace{-0.4cm}
\end{figure}

\begin{figure}[t]
\center
				\includegraphics[width=0.37\textwidth]{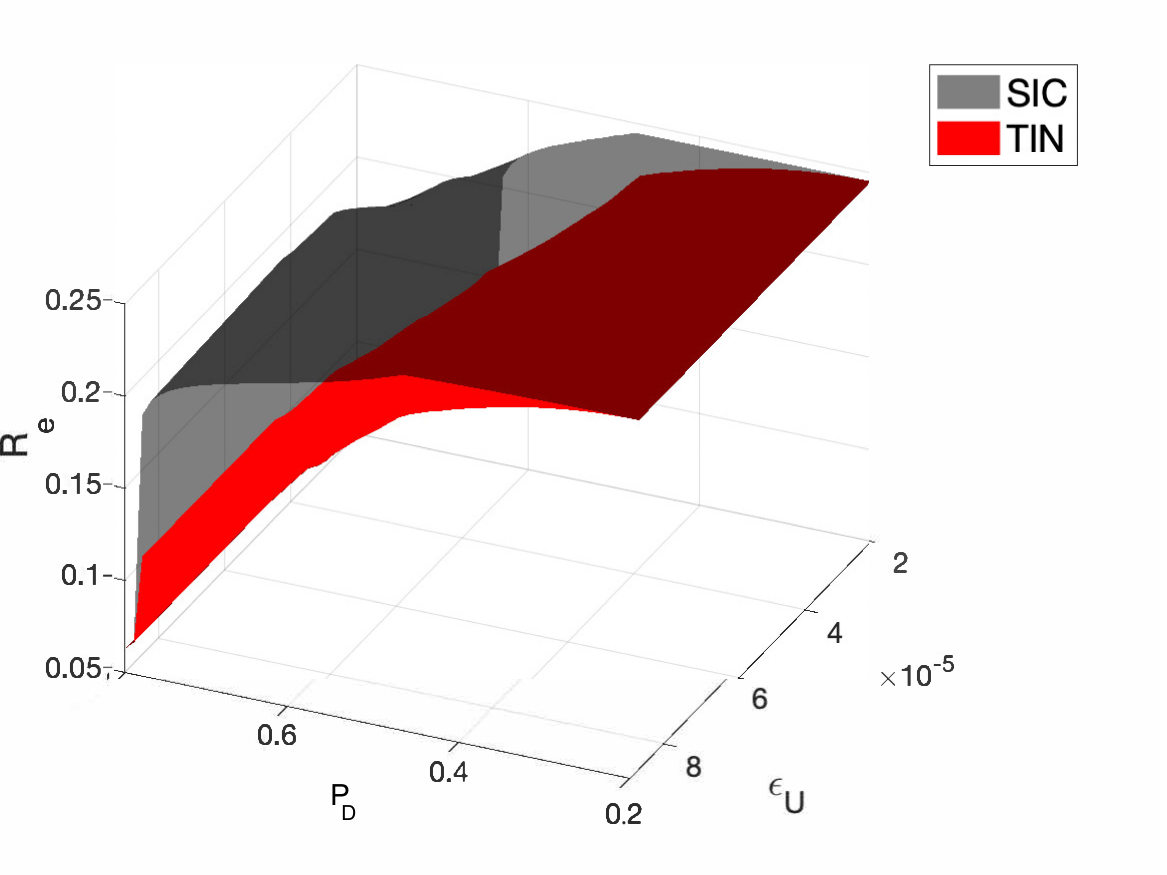}
\caption{An illustration of the upper bound on the eMBB transmission as a function of $P_{\D}$ and $\epsilon_{\U}$ under the TIN and SIC approaches. In this figure, $\epsilon_{\U}$ varies from $10^{-5}$ to $10^{-4}$ with a step size of $10^{-5}$. In this figure, $\P = 0.5$, $\ell = 150$, $\eta = 10$, $P_{\text{FA}} = 10^{-4}$ and $\epsilon_{\e} = 10^{-2}$.}
\label{fig5}
\vspace{-0.4cm}
\end{figure}


\section{Numerical Analysis} \label{sec:analysis}
In this section, we numerically evaluate the rate-reliability-detection trade-off of Theorem~\ref{th1} and Theorem~\ref{th2}. For given $\P, n, \eta, P_{\text{FA}}, \epsilon_{\e}$ and $\epsilon_{\U}$, we first find values of $\alc$ and $\bc$ such that $\epsilon_{b, \U}$ is below  $\epsilon_{\U}$  in all blocks. Meaning that the constraint \eqref{eq:urllc} is satisfied for all blocks. Next, we maximize the upper bound on the eMBB rate  in \eqref{eq:embb} over $\bsf, \bss, \alsf, \alss$ while satisfying the minimum required target detection probability (i.e., $P_{\D}$) over all blocks.

We compare our results with the following two baseline schemes: 
\begin{itemize}
\item \emph{Power-sharing scheme}: under this scheme, sensing and communication signals are superposed by simply adding the two corresponding codewords \cite{Li2023Globecom}.  We can obtain the power-sharing scheme from the proposed DPC-based scheme by  setting 
$
\alc = \alsf = \alss = 0.
$
We then share the total transmit power $\P$ between the communication and sensing tasks using the power sharing parameters $\bsf, \bc$ and $\bss$ as explained in our scheme. 
\item \emph{Time-sharing scheme}: under this scheme, the available transmission block is shared between the sensing and communication tasks and thus the two tasks are performed separately \cite{Keshtiarast2025}. To analyze the performance of this scheme,  in each block~$b \notin \Barr$, we share the $\ell$ channel uses between the communication task of transmitting the eMBB message and the sensing task. In each block~$b \in \Barr$, we share the $\ell$ channel uses among the communication task of transmitting a URLLC message, transmitting the eMBB message and the sensing task. Each individual task is performed using the transmit power $\P$. 

\end{itemize}


 Fig.~\ref{fig2} illustrates the upper bound on the eMBB transmission rate as a function of $P_{\D}$ for  our DPC-based scheme under the TIN approach. We compare the result with the power-sharing and  time-sharing schemes when $\epsilon_{\U}$ is fixed at $10^{-2}$. As can be seen from this figure,  our scheme significantly outperforms the other two schemes. 
 Fig.~\ref{fig3} illustrates the upper bound on the eMBB transmission rate as a function of $P_{\D}$ for our  proposed TIN scheme under different  levels of  URLLC reliability. It can be seen that as we decrease the required threshold on the URLLC decoding error probability (i.e., $\epsilon_{\U}$) the eMBB rate decreases. In other words, increasing URLLC reliability requirements decreases the eMBB transmission rate.  
 
 Next we compare the performance of our DPC-based scheme under the TIN and the SIC approaches. 
 In Fig~\ref{fig4}, we plot the upper bound on the eMBB transmission rate $R_{\e}$ under both approaches for different values of $\epsilon_{\U}$. It can be seen that in the case of no URLLC or when the URLLC reliability constraint is very low, i.e., $\epsilon_{\U} = 10^{-2}$, the TIN and the SIC approaches have similar performance. However, at high URLLC requirements, the SIC approach outperforms the TIN.   
 Fig.~\ref{fig5} illustrates the upper bound on the eMBB transmission as a function of $P_{\D}$ and $\epsilon_{\U}$ for both the TIN and  SIC approaches. In this plot, $\epsilon_{\U}$ varies from $10^{-5}$ to $10^{-4}$ with a step size of $10^{-5}$ which implies high requirements on the URLLC reliability and results in the outperformance of the SIC over the TIN.  
 However, the performance gap decreases when the target detection probability, i.e., $P_{\D}$, is either very low or very high. Specifically, when the target detection probability is very low,  most blocks are with no URLLC, resulting in comparable performance for both approaches. At high target detection probability, most blocks are with URLLC and thus more subtracted URLLC interference will be wrong which introduces error in the eMBB decoding under the SIC and eventually reduces the gap between the two approaches.

\section{Conclusion} \label{sec:conclusion}
We have proposed a MIMO  ISAC-enabled URLLC system where a BS communicates with a UE and a SR collects echo signals reflected from a target of interest.  The BS simultaneously transmit messages from eMBB and URLLC services. During each eMBB transmission, the transmission of an additional URLLC message is triggered when the SR sensed the presence of the target. To reinforce URLLC transmissions, the interference of both eMBB and sensing signals   were mitigated using DPC. For this system, we  formulated the rate-reliability-detection trade-off in the finite blocklength regime.  Our numerical analysis shows a significant outperformance of our scheme over  the power-sharing  and  the  time-sharing methods. 
\appendices 
\section{Proof of Lemma~\ref{lemma2}}\label{App:B}
Recall that all the codewords are uniformly distributed on the power shell. As a result of this code construction, our input codewords are non i.i.d.
We thus take a change of metric approach and instead of the information density metric $i^{(\U)}_b  (\vectsf V_b; \vectsf Y_{b,c} )$ defined in \eqref{eq:ibu}, we use the following metric: 
\begin{IEEEeqnarray}{rCl}\label{eq:iu}
\tilde i^{(\U)}_b  (\vectsf V_b; \vectsf Y_{b,c} ) 
&=& \sum_{j = 1}^q \log \frac{Q_{\vect Y_{b,c,j}| \vect V_{b,j}} (\vectsf y_{b,c}| \vectsf v_{b})}{Q_{\vectsf Y_{b,c}}(\vectsf y_{b,c})} ,
\end{IEEEeqnarray}
where $Q$'s are i.i.d Gaussian distributions. 
Specifically, 
\begin{IEEEeqnarray}{rCl}
Q_{\vect Y_{b,c,j}}& \sim & \mathcal N \left (\vect y_{b,c,j}: \vect 0, \sigy \vectsf I_{\ell \times \ell} \right), \label{eq:qy} \\
Q_{\vect Y_{b,c,j}| \vect V_{b,j}} &\sim& \mathcal N \left (\vect y_{b,c,j}:\sqrt{\lambda_{b,j}} \vect v_{b,j},  \sigyvj \vectsf I_{\ell \times \ell} \right), \IEEEeqnarraynumspace \label{eq:qyv}
\end{IEEEeqnarray}
where $\sigy$ and $\sigyvj$ are defined in \eqref{eq:sigy} and \eqref{eq:sigyvj}, respectively. 
The following lemma shows that the random variable $\tilde i^{(\U)}_b  (\vectsf V_b; \vectsf Y_{b,c} )$ converges in distribution to a Gaussian distribution. 
\begin{lemma}\label{lemma5}
The following holds: 
\begin{IEEEeqnarray}{rCl}
\tilde i^{(\U)}_b  (\vectsf V_b; \vectsf Y_{b,c} ) \sim \mathcal N(\ell \C_{\U},\ell \mathsf V_{\U} ),
\end{IEEEeqnarray}
where 
\begin{IEEEeqnarray}{rCl}
\C_{\U} : =  \sum_{j = 1}^q \C(\Omega_{b,j}),\quad 
\mathsf V_{\U} := \sum_{j = 1}^q \mathsf V(\Omega_{b,j})
\end{IEEEeqnarray}
with $\C(x) = \frac{1}{2} \log (1+x)$, $\mathsf V(x): = \frac{x(2+x)}{2(1+x)^2}$ and $\Omega_{b,j}$ is defined in \eqref{eq:omegabj}. 
\begin{IEEEproof}
See Appendix~\ref{App:E}. 
\end{IEEEproof}
\end{lemma}
We are now ready to analyze the URLLC decoding error probability. Recall the bound on the average URLLC decoding error probability from \eqref{eq:57}.  
Recall the definition of $\mathcal E_{\U,1}$, $\mathcal E_{\U,2}$ and $\mathcal E_{\U,3}$ from \eqref{eq:eu1}, \eqref{eq:eu2} and \eqref{eq:eu3}, respectively. 
\subsubsection{Analyzing $\Pr [\mathcal E_{\U,1}]$} This is equivalent to the probability that for all $v \in [ M_{v}]$ and for all $m \in [M_{\U}]$ there is no codeword $ \vectsf V_b(m,v)$ such that $\tilde i^{(\U)}_b  (\vectsf V_b; \vectsf Y_{b,c} ) > \delta_{\U}$. Hence, 
\begin{eqnarray}
\Pr [\mathcal E_{\U,1}] = \left (\Pr [ \tilde i^{(\U)}_b  (\vectsf V_b; \vectsf Y_{b,c} ) \le \delta_{\U} ] \right)^{M_{\U} M_{v}}. \label{eq:58}
\end{eqnarray}
\subsubsection{Analyzing $\Pr [\mathcal E_{\U,2}]$}
This is equivalent to the probability that the UE has detected a URLLC message but the decoded one does not match the transmitted one.  To evaluate this probability, we use the threshold bound for maximum-metric decoding. I.e., for any given threshold $\delta_{\U}$: 
\begin{IEEEeqnarray}{rCl}
\Pr [\mathcal E_{\U,2}] &\le& \Pr [\tilde i^{(\U)}_b  (\vectsf V_b(m_{b,\U}, v); \vectsf Y_{b,c} ) \le \delta_{\U} ] \notag \\
&& + (M_{\U} M_v -1)  \Pr [\tilde i^{(\U)}_b  (\bar {\vectsf V}_b(m', v'); \vectsf Y_{b,c} ) > \delta_{\U} ] \label{eq:59} \IEEEeqnarraynumspace
\end{IEEEeqnarray}
where $m' \in [M_{\U}]$, $v' \in [M_v]$ and $(m_{b,\U}, v) \neq (m', v')$ and $\bar {\vectsf V}_b \sim f_{\vectsf V_b}$ and is independent of $\vectsf V_b$ and $ \vectsf Y_{b,c} $. 

\subsubsection{Analyzing $\Pr [\mathcal E_{\U,3}]$} This is equivalent to the probability that no URLLC message has been sent over block~$b$ but there is at least one codeword $\bar {\vectsf V}_b$ such that $\Pr [\tilde i^{(\U)}_b  (\bar{\vectsf V}_b; \vectsf Y_{b,c} ) > \delta_{\U}]$.  Hence,  
\begin{eqnarray}
\Pr [\mathcal E_{\U,3}] = 1 - \left (\Pr [\tilde i^{(\U)}_b  (\bar{\vectsf V}_b; \vectsf Y_{b,c} ) \le \delta_{\U}]\right )^{M_{\U} M_{v}}. \label{eq:60}
\end{eqnarray}
Combining \eqref{eq:57}, \eqref{eq:58}, \eqref{eq:59} and \eqref{eq:60}, we have
\begin{IEEEeqnarray}{rCl}
\epsilon_{b,\U}
&\le& P_{b-1, \D} \Big( \Pr [ \tilde i^{(\U)}_b  (\vectsf V_b; \vectsf Y_{b,c} ) \le \delta_{\U} ]^{M_{\U} M_{v}} \notag \\
&& \hspace{1.25cm}+ \Pr [\tilde i^{(\U)}_b  (\vectsf V_b; \vectsf Y_{b,c} ) \le \delta_{\U} ] \notag \\
&&\hspace{1.5cm}+ (M_{\U} M_v -1)  \Pr [\tilde i^{(\U)}_b  (\bar {\vectsf V}_b; \vectsf Y_{b,c} ) > \delta_{\U} ]\Big)\notag \\
&&\hspace{-0.7cm}+ (1- P_{b-1, \D} )\left(1 - \left (\Pr [\tilde i^{(\U)}_b  (\bar{\vectsf V}_b; \vectsf Y_{b,c} ) \le \delta_{\U}]\right )^{M_{\U} M_{v}}\right) . \label{eq:euf}\IEEEeqnarraynumspace
\end{IEEEeqnarray}

For some $K_{\U} > 0$, set 
\begin{eqnarray}
\delta_{\U} := \log(M_{\U}M_v)+ K_{\U} \log (\ell).
\end{eqnarray}
To evaluate  $\Pr [i^{(\U)}_b  (\bar {\vectsf V}_b; \vectsf Y_{b,c} ) > \delta_{\U} ]$ we follow \cite[Lemma 47]{Yuri2012} which results in 
\begin{eqnarray} \label{eq:98}
\Pr [\tilde i^{(\U)}_b  (\bar {\vectsf V}_b; \vectsf Y_{b,c} ) > \delta_{\U} ] \le \frac{2}{M_{\U} M_{v} \ell^{K_{\U} }} \left ( \frac{\log 2}{\sqrt{2\pi\ell}} + \frac{2B} {\sqrt{\ell}}\right).
\end{eqnarray}

Given that $\tilde i^{(\U)}_b  (\vectsf V_b(m_{b,\U}, v); \vectsf Y_{b,c} )$ follows a Gaussian distribution (as shown in Lemma~\ref{lemma5}), thus  we employ  the Berry-Esseen CLT for functions \cite[Proposition~1]{MolavianJaziArXiv} to bound the following probability:
\begin{IEEEeqnarray}{rCl} \label{eq:99}
\lefteqn{\Pr [\tilde i^{(\U)}_b  (\vectsf V_b; \vectsf Y_{b,c} ) \le \delta_{\U}  ]} \notag \\
& \le& Q \left (\frac{-\log(M_{\U} M_v) + \ell \C_{\U} - K_{\U} \log (\ell)}{\sqrt{\ell \mathsf V_{\U}}} \right) + \frac{B}{\sqrt{\ell} }, \IEEEeqnarraynumspace
\end{IEEEeqnarray}
where $B > 0$ is a constant.
Combining \eqref{eq:98} and \eqref{eq:99} with \eqref{eq:euf} concludes the proof. 

\section{Proof of Lemma~\ref{lemma3}} \label{App:C}
Recall the average decoding error probability of eMBB under the TIN approach from \eqref{eq:51}. 
We now evaluate the right-hand side (RHS) of \eqref{eq:51}. 
\subsubsection{Analyzing $\Pr [\Bdet = B_{\text{dt}}] $}
\begin{IEEEeqnarray}{rCl} \label{eq:Pdt}
\tilde \P_{\text{dt}} &: =& \Pr [b \in \Barr] \Pr [b \in \Bdet | b \in \Barr] \notag \\
&& + \Pr [b \notin \Barr] \Pr [b \in \Bdet | b \notin \Barr] \\
& = & \P_{b-1,\D} (1- \Pr [\mathcal E_{\U,1}]) +  (1-\P_{b-1,\D}) \Pr [\mathcal E_{\U,3}]. 
\end{IEEEeqnarray}
We thus have
\begin{eqnarray}
\Pr [\Bdet = B_{\text{dt}}]  = \tilde \P_{\text{dt}}^{|\Bdt|} (1- \tilde \P_{\text{dt}} )^{\eta-|\Bdt|}, \label{eq:74}
\end{eqnarray}
which is equivalent to the probability that the UE detects URLLC messages in all the blocks in $\Bdt$ and there  is no URLLC detection in blocks that are in $[\eta] \backslash \Bdt$. 
\subsubsection{Analyzing $\Pr [\mathcal E_{\e,1}| \Bdet =  B_{\text{dt}} ] $}
This error event is equivalent to the probability that the set of blocks in which UE detects URLLC is different  from the set of blocks in which the BS transmits  URLLC messages (i.e., $\Bdet \neq \Barr$).  
%
\begin{IEEEeqnarray}{rCl}
\lefteqn{\Pr [\mathcal E_{\e,1} | \Bdet =  B_{\text{dt}}]} \notag \\
& = & \Pr [ \Barr \neq B_{\text{dt}} | \Bdet = B_{\text{dt}} ] \\
& = & 1 - \frac{\Pr [\Barr =  B_{\text{dt}}] \Pr [ \Bdet =  B_{\text{dt}}| \Barr = B_{\text{dt}}  ]}{\Pr[\Bdet  = B_{\text{dt}}]}\IEEEeqnarraynumspace \\
& = & 1 - \frac{(P_{b-1, \D}(1- \Pr [\mathcal E_{\U,1}]))^{|\Bdt|}  }{ \tilde \P_{\text{dt}}^{|\Bdt|} }.\notag \\
&&\hspace{1cm} \times \frac{ ((1-\Pr [\mathcal E_{\U,3}])(1- P_{b-1, \D}))^{\eta - |\Bdt|} }{(1- \tilde \P_{\text{dt}} )^{\eta-|\Bdt|}}\label{eq:79}
\end{IEEEeqnarray}

\subsubsection{Analyzing $\Pr [\mathcal E_{\e,2}|\Bdet =  B_{\text{dt}} ]$}
 To evaluate this probability, we first follow the argument provided in Appendix~\ref{App:B}, and introduce a new metric:
 \begin{IEEEeqnarray}{rCl}
\lefteqn{\tilde i^{(\e)}_{\text{TIN}} \left ( \vectsf s_{\e,2} ,  \vectsf s_{\e,1} ;  \vectsf y_c| \mathcal B_{\text{detect}} = B_{\text{dt}} \right)}\notag \\ 
& = & \sum_{b\notin  B_{\text{dt}}  }\sum_{j = 1}^q \log \hspace{0cm}  \frac{Q_{\vect Y_{b,c,j}| \vect S_{b,j}^{(1)}} (\vect y_{b,c,j}| \vect s_{b,j}^{(1)})}{Q_{\vect Y_{b,c.j}}(\vect y_{b,c,j})} \notag \\
&& + \sum_{b\in  B_{\text{dt}} }  \sum_{j = 1}^q \log \hspace{0cm}  \frac{Q_{\vect Y_{b,c,j} | \vect S_{b,j}^{(2)}} (\vect y_{b,c,j}| \vect s_{b,j}^{(2)})}{Q_{\vect Y_{b,c,j}}(\vect y_{b,c,j})},
\end{IEEEeqnarray}
where $Q$s are Gaussian distributions. Specifically, $Q_{\vect Y_{b,c,j}}( \vect y_{b,c,j})$ is defined in \eqref{eq:qy} and 
\begin{IEEEeqnarray}{rCl}
Q_{\vect Y_{b,c,j}| \vect S_{b,j}^{(1)}}  &\sim& \mathcal N \left (\vect y_{b,c,j}:\lambda_{b,j} \vect s_{b,j}^{(1)}, \sigysfj \vectsf I_{\ell \times \ell}   \right),\label{eq:qysf}\\
Q_{\vect Y_{b,c,j}| \vect S_{b,j}^{(2)}} &\sim& \mathcal N \left (\vect y_{b,c,j}:\lambda_{b,j} (1-\alc)^2\vect s_{b,j}^{(2)}, \sigyssj \vectsf I_{\ell \times \ell}   \right),\IEEEeqnarraynumspace \label{eq:qyss}
\end{IEEEeqnarray}
where $\sigysfj$ and $\sigyssj$ are defined in \eqref{eq:sigysfj} and \eqref{eq:sigyssj}, respectively.   
Following the same argument as the proof of  Lemma~\ref{lemma5} in \cite[Appendix~E]{arXiv}, we have  the following lemma showing $\tilde i^{(\e)}_{\text{TIN}} \left ( \vectsf s_{\e,2} ,  \vectsf s_{\e,1} ;  \vectsf y_c| \mathcal B_{\text{detect}} = B_{\text{dt}} \right)$ converges in distribution to a Gaussian distribution. 
\begin{lemma} \label{lemma6}
The following holds:
\begin{IEEEeqnarray}{rCl}
\tilde i^{(\e)}_{\text{TIN}} \left ( \vectsf s_{\e,2} ,  \vectsf s_{\e,1} ;  \vectsf y_c| \mathcal B_{\text{detect}} = B_{\text{dt}} \right) \sim  \mathcal N(n \tilde{\C}_{\e},n \tilde{\mathsf V}_{\e}), 
\end{IEEEeqnarray}
where $\tilde{\C}_{\e}$ and $\tilde{\mathsf V}_{\e}$ are defined in \eqref{eq:tce} and \eqref{eq:tve}, respectively. 
\end{lemma}
Now by using the threshold bound for maximum-metric decoding, for any given threshold $\delta_{\e}$, we have 
\begin{IEEEeqnarray}{rCl} \label{eq:130}
\lefteqn{\Pr [\mathcal E_{\e,2}|\Bdet = B_{\text{dt}} ]} \notag \\ 
&\le& \Pr [\tilde i^{(\e)}_{\text{TIN}} \left ( \vectsf S_{\e,1},  \vectsf S_{\e,2};  \vectsf Y_c| \mathcal B_{\text{detect}} = B_{\text{dt}} \right) \le \delta_{\e} ] \notag \\
&+& (M_{\e}M_{\s} -1)  \Pr [\tilde i^{(\e)}_{\text{TIN}} \left ( \bar{ \vectsf S}_{\e,1},  \bar{ \vectsf S}_{\e,2};  \vectsf Y_c| \mathcal B_{\text{detect}} = B_{\text{dt}} \right) > \delta_{\e} ], \IEEEeqnarraynumspace \label{eq:80}
\end{IEEEeqnarray}
where $\bar{ \vectsf S}_{\e,1} \sim f_{ \vectsf S_{\e,1}}$ and $\bar{ \vectsf S}_{\e,2} \sim f_{ \vectsf S_{\e,2}}$ and are independent of $ \vectsf S_{\e,1}$, $ \vectsf S_{\e,2}$ and $ \vectsf Y_{c} $. 
For some $K_{\e} > 0$, set
\begin{eqnarray}
\delta_{\e} := \log M_{\e} + \log M_{\s} + K_{\e} \log (n).
\end{eqnarray}

To evaluate $\Pr [\tilde i^{(\e)}_{\text{TIN}} \left (  \vectsf S_{\e,1},  \vectsf S_{\e,2};  \vectsf Y_c| \mathcal B_{\text{detect}} = B_{\text{dt}} \right) \le \delta_{\e} ]$, given that $\tilde i^{(\e)}_{\text{TIN}} \left (  \vectsf S_{\e,1},  \vectsf S_{\e,2};  \vectsf Y_c| \mathcal B_{\text{detect}} = B_{\text{dt}} \right) $ follows a Gaussian distribution (see Lemma~\ref{lemma6}),  we employ  the Berry-Esseen CLT for functions to bound this probability. Hence,
\begin{IEEEeqnarray}{rCl}
\lefteqn{\Pr [\tilde i^{(\e)}_{\text{TIN}} \left (  \vectsf S_{\e,1},   \vectsf S_{\e,2};  \vectsf Y_c| \mathcal B_{\text{detect}} = B_{\text{dt}} \right) \le \delta_{\e} ]} \notag \\
& \le& Q\left (\frac{-\log(M_{\e}) - \log(M_{\s}) + n \tilde {\C}_{\e} - K_{\e} \log (n)}{\sqrt{n \tilde {\mathsf V}_{\e}}} \right) + \frac{\tilde B}{\sqrt{n } }, \notag \\\label{eq:82}
\end{IEEEeqnarray}
for some $\tilde B>0$ and where $Q (\cdot)$ is the Q-function. To evaluate  $\Pr [\tilde i^{(\e)}_{\text{TIN}} \left ( \bar{ \vectsf S}_{\e,1},  \bar{ \vectsf S}_{\e,2};  \vectsf Y_c| \mathcal B_{\text{detect}} = B_{\text{dt}} \right) > \delta_{\e} ]$ we follow \cite[Lemma 47]{Yuri2012} which results in 
\begin{IEEEeqnarray}{rCl}
\lefteqn{\Pr [\tilde i^{(\e)}_{\text{TIN}} \left ( \bar{ \vectsf S}_{\e,1},  \bar{ \vectsf S}_{\e,2};  \vectsf Y_c| \mathcal B_{\text{detect}}= B_{\text{dt}} \right) > \delta_{\e}]}\notag \\
& \le& \frac{2}{M_{\e}M_{\s} n^{K_{\e} }} \left ( \frac{\log 2}{\sqrt{2\pi n}} + \frac{2 \tilde B} {\sqrt{n}}\right). \label{eq:83}
\end{IEEEeqnarray}

By combining \eqref{eq:71}, \eqref{eq:74}, \eqref{eq:79}, \eqref{eq:80}, \eqref{eq:82} and \eqref{eq:83} we have the following bound:
\begin{IEEEeqnarray}{rCl}
\epsilon_{\e}^{(n)}& \stackrel{(i)}{\le} & 1 + \frac{\tilde B}{\sqrt{n}}\left ( 1 + \frac{4}{n^{K_{\e}}}\right)  +\frac{2\log 2}{n^{K_{\e}}\sqrt{2n\pi}}\notag \\
&&-\sum_{\Bdt} \left (P_{b-1, \D}(1-(\tepf)^{M_{\U}M_v})\right)^{|\Bdt|}\notag \\
&& \hspace{1cm} \times \left ((1-P_{b-1, \D}) (1-P_{\U,2}) \right)^{\eta - |\Bdt|}\notag \\
&& + \sum_{\Bdt}  \P_{\text{dt}}^{|\Bdt|} (1-\P_{\text{dt}})^{\eta-|\Bdt|} \notag \\
&&\hspace{0.5cm}Q\left (\frac{-\log(M_{\e}M_{\s}) + n \tilde {\C}_{\e} - K_{\e} \log (n)}{\sqrt{n \tilde {\mathsf V}_{\e}}} \right) \\
& = & \Delta_{\e} + \sum_{\Bdt}  \P_{\text{dt}}^{|\Bdt|} (1-\P_{\text{dt}})^{\eta-|\Bdt|} \notag \\
&&\hspace{0.5cm}Q\left (\frac{-\log(M_{\e}M_{\s}) + n \tilde {\C}_{\e} - K_{\e} \log (n)}{\sqrt{n \tilde {\mathsf V}_{\e}}} \right), \label{eq:136}
\end{IEEEeqnarray}

where $\Delta_{\e}$ is defined in \eqref{eq:Deltae} and $(i)$ is followed by \eqref{eq:58} and \eqref{eq:60} which also result in  $\tilde \P_{\text{dt}} \le \P_{\text{dt}}$ where $\P_{\text{dt}}$ is defined in \eqref{eq:Pdtn}. 
According to Proposition~\ref{prop1}, it is required that $\epsilon_{\e}^{(n)}$ to be bounded above by $ \epsilon_{\e}$, i.e., $\epsilon_{\e}^{(n)} \le \epsilon_{\e}$. Thus
 \begin{IEEEeqnarray}{rCl}
\lefteqn{\epsilon_{\e} \ge \Delta_{\e} + \sum_{\Bdt}  \P_{\text{dt}}^{|\Bdt|} (1-\P_{\text{dt}})^{\eta-|\Bdt|}}\notag \\
&&\hspace{1.5cm} Q\left (\frac{-\log(M_{\e}M_{\s}) + n \tilde {\C}_{\e} - K_{\e} \log (n)}{\sqrt{n \tilde {\mathsf V}_{\e}}} \right) \IEEEeqnarraynumspace\\
&\stackrel{(i)}{\ge}& \Delta_{\e} + Q\left (\frac{-\log(M_{\e}M_{\s}) + n  {\C}_{\e} - K_{\e} \log (n)}{\sqrt{n  {\mathsf V}_{\e}}} \right)
\end{IEEEeqnarray}
where $\C_{\e}$ and $\mathsf V_{\e}$ are defined in \eqref{eq:Ce} and \eqref{eq:Ve}, respectively, and $(i)$ follows by the fact that $Q$-function is a decreasing function of its argument. 
By taking the inverse of the $Q$-function and dividing both sides  by $n$, we conclude the proof. 

\section{Proof of Lemma~\ref{lemma4}} \label{App:D}
Recall the average decoding eMBB error probability under the SIC approach from \eqref{eq:61}. In this section, we analyze $\Pr[\Bdec = \Bdc|\Bdet = \Barr, \Bdet = \Bdt]$ and $\Pr [\mathcal E_{\e,2} | \Bdet = \Bdt , \Bdec = \Bdc]$. See Appendix~\ref{App:C} for the calculation of the remaining terms in the RHS of \eqref{eq:61}.  
\subsection{Analyzing $\Pr[\Bdec = \Bdc|\Bdet = \Barr, \Bdet = \Bdt]$}
From \eqref{eq:59}, we have 
\begin{IEEEeqnarray}{rCl} \label{eq:Pdc}
 \Pr [m_{b, \U} \neq \hat m_{b,\U} | \Bdet = \Barr = \Bdt ] 
 \le \tepf + \teps. \IEEEeqnarraynumspace
\end{IEEEeqnarray}
Let $\tepf + \teps: =1 - \P_{\text{dc}}$.
Thus, 
\begin{IEEEeqnarray}{rCl}
\lefteqn{\Pr[\Bdec = \Bdc|\Bdet = \Barr, \Bdet = \Bdt]} \notag \\
& = & \prod_{b \in \Bdc} \Pr [m_{b, \U} = \hat m_{b,\U} | \Bdet = \Barr = \Bdt ] \notag \\
&& \cdot \prod_{ b \in \Bdt \backslash \Bdc} (1-\Pr [m_{b, \U} = \hat m_{b,\U} | \Bdet = \Barr = \Bdt ] ) \IEEEeqnarraynumspace \\
& \le & \P_{\text{dc}}^{|\Bdc|} (1- \P_{\text{dc}})^{|\Bdt| - |\Bdc|}.
\end{IEEEeqnarray}
which is equivalent to the probability that all the URLLC messages in $\Bdc$ are decoded correctly and no URLLC message is decoded correctly in the remaining blocks in $\Bdt \backslash \Bdc$. 
\subsection{Analyzing $\Pr [\mathcal E_{\e,2} | \Bdet = \Bdt , \Bdec = \Bdc]$}
 To evaluate this probability, we first follow the argument provided in Appendix~\ref{App:B}, and introduce a new metric:
 \begin{IEEEeqnarray}{rCl}
\lefteqn{\tilde i^{(\e)}_{\text{SIC}} \left ( \vectsf s_{\e,2} ,  \vectsf s_{\e,1} ;  \vectsf y_c| \Bdet = \Bdt, \Bdec = \Bdc \right):=}\notag \\ &&
 \log \hspace{-0.15cm}\prod_{b\notin  \Bdt  }\hspace{0cm}  \frac{Q_{\vectsf Y_{b,c}| \vectsf S_{b}^{(1)}} (\vectsf y_{b,c}| \vectsf s_{b}^{(1)})}{Q_{\vectsf Y_{b,c}}(\vectsf y_{b,c})} +  \log \hspace{-0.15cm}\prod_{b\in  \Bdt \backslash \Bdc } \hspace{0cm}  \frac{Q_{\vectsf Y_{b,c} | \vectsf S_{b}^{(2)}} (\vectsf y_{b,c}| \vectsf s_{b}^{(2)})}{Q_{\vectsf Y_{b,c}}(\vectsf y_{b,c})} \notag \\
 && +  \log \hspace{-0.15cm}\prod_{b\in \Bdc } \hspace{0cm}  \frac{Q_{\vectsf Y_{b,c} | \vectsf S_{b}^{(2)}, \vectsf V_b} (\vectsf y_{b,c}| \vectsf s_{b}^{(2)}, \vectsf v_b)}{Q_{\vectsf Y_{b,c}| \vectsf V_b}(\vectsf y_{b,c}|\vectsf v_b)},\IEEEeqnarraynumspace
\end{IEEEeqnarray}
where $Q$s are Gaussian distributions. Specifically, $Q_{\vect Y_{b,c,j}}( \vect y_{b,c,j})$, $Q_{\vectsf Y_{b,c}| \vectsf V_b}(\vectsf y_{b,c}|\vectsf v_b)$, $Q_{\vectsf Y_{b,c}| \vectsf S_{b}^{(1)}} (\vectsf y_{b,c}| \vectsf s_{b}^{(1)})$, and $Q_{\vectsf Y_{b,c} | \vectsf S_{b}^{(2)}} (\vectsf y_{b,c}| \vectsf s_{b}^{(2)})$ are defined in \eqref{eq:qy}, \eqref{eq:qyv}, \eqref{eq:qysf} and \eqref{eq:qyss}, respectively, and $Q_{\vectsf Y_{b,c} | \vectsf S_{b}^{(2)}, \vectsf V_b} (\vectsf y_{b,c}| \vectsf s_{b}^{(2)}, \vectsf v_b)$ is a Gaussian distribution with mean $\vbj + (1-\alc) \sbjf$ and variance $\sigysvj$ defined in \eqref{eq:sigysvj}. 
Following the same argument as the proof of Lemma~\ref{lemma5}, we have  the following lemma.
\begin{lemma}
The following holds:
\begin{equation}
\hspace{-0.2cm}\tilde i^{(\e)}_{\text{SIC}} \left ( \vectsf s_{\e,2} ,  \vectsf s_{\e,1} ;  \vectsf y_c|  \Bdet = \Bdt, \Bdec = \Bdc\right) \sim  \mathcal N(n \tilde{\C}_{\e,2},n \tilde{\mathsf V}_{\e,2}), 
\end{equation}
where $\tilde{\C}_{\e,2}$ and $\tilde{\mathsf V}_{\e,2}$ are defined in \eqref{eq:tce2} and \eqref{eq:tve2}, respectively. 
\end{lemma}
Following similar steps as in \eqref{eq:130} to \eqref{eq:136}, we can
show that 
\begin{IEEEeqnarray}{rCl}
\lefteqn{\epsilon_{\e} \ge \Delta_{\e}}\notag \\
&& + \sum_{\Bdt}  \P_{\text{dt}}^{|\Bdt|} (1-\P_{\text{dt}})^{\eta-|\Bdt|} \sum_{\Bdc}  \P_{\text{dc}}^{|\Bdc|} (1- \P_{\text{dc}})^{|\Bdt| - |\Bdc|}\notag \\
&&\hspace{1.5cm} Q\left (\frac{-\log(M_{\e}M_{\s}) + n \tilde {\C}_{\e,2} - K_{\e} \log (n)}{\sqrt{n \tilde {\mathsf V}_{\e,2}}} \right) \IEEEeqnarraynumspace\\
&\ge& \Delta_{\e} + Q\left (\frac{-\log(M_{\e}M_{\s}) + n  {\C}_{\e,2} - K_{\e,2} \log (n)}{\sqrt{n  {\mathsf V}_{\e,2}}} \right)
\end{IEEEeqnarray}
where $\C_{\e,2}$ and  $\mathsf V_{\e,2}$ are defined in \eqref{eq:Ce2} and  \eqref{eq:Ve2}, respectively. By taking the inverse of the $Q$-function and dividing both sides  by $n$, we conclude the proof.

\section{Proof of Lemma~\ref{lemma1}} \label{App:A}
From \eqref{eq:test3}, the probability of false alarm  is given by 
\begin{IEEEeqnarray}{rCl}
P_{b, \text{FA}} 
& = & \Pr \left [ \sum_{j = 1}^q \left (|| \vect N_{b,s,j}||^2 - \frac{||\vect N_{b,s,j} -  \vect \mu_{b,j}  ||^2}{\sigbj} \right) > \delta \right ] \IEEEeqnarraynumspace\\
& = & \Pr \left [ u_b  > \delta +  \sum_{j = 1}^q\frac{1}{\sigbj-1}||\vect \mu_{b,j}||^2   \right ],
\end{IEEEeqnarray}
where 
\begin{equation}
u_b := \sum_{j = 1}^q \left (1-\frac{1}{\sigbj}\right) ||\vect N_{b,s,j}+ \frac{1}{\sigbj -1}\vect \mu_{b,j} ||^2.
\end{equation}
Given that $\vect N_{b,s,j} \sim  \mathcal N(\vect 0, \I_{\ell})$, then   $\vect N_{b,s,j} + \frac{1}{\sigbj -1}\vect \mu_{b,j} \sim \mathcal N( \frac{1}{\sigbj -1}\vect \mu_{b,j}, \vectsf I_{\ell})$ and consequently $||\vect N_{b,s,j}+ \frac{1}{\sigbj -1}\vect \mu_{b,j} ||^2
$ follows a non-central chi-square distribution with degree of $\ell$ and parameter $\frac{1}{(\sigbj -1)^2}||\vect \mu_{b,j}||^2$ . Hence, $u_{b}$ is the weighted sum of independent non-central chi-square random variables. It thus follows a generalized chi-square distribution. More specifically, 
\begin{equation}
u_b \sim \tilde {\mathcal X}_1 (\{ w_{b,j}\}_{j =1}^{q}, \ell, \{ \nu_{b,j}\}_{j = 1}^q),
\end{equation}
where the parameters $\{ w_{b,j}\}_{j =1}^{q}$ and $\{ \nu_{b,j}\}_{j = 1}^q$ are defined in \eqref{eq:27}. 
Let $F_{\tilde{\mathcal X}_1} (\cdot)$ denote the CDF of the corresponding generalized chi-square distribution.  The probability of false alarm thus is given by
\begin{eqnarray}
P_{b, \text{FA}} = 1 - F_{\tilde{\mathcal X}_1}\left (\delta +  \sum_{j = 1}^q\frac{1}{\sigbj-1}||\vect \mu_{b,j}||^2\right).
\end{eqnarray}
For each block~$b$, we fix $P_{b, \text{FA}}$ at $P_{\text{FA}}$. Thus, $\delta$ is equal to
\begin{eqnarray}
\delta = F^{-1}_{\tilde{\mathcal X}_1} (1-P_{\text{FA}})-\sum_{j = 1}^q\frac{1}{\sigbj-1}||\vect \mu_{b,j}||^2.
\end{eqnarray} 
The target detection probability is given by
\begin{IEEEeqnarray}{rCl}
P_{b, \D} 
& = & \Pr \Big [ \sum_{j = 1}^q \Big (|| \vect N_{b,s,j}  +  \vect X_{b,j} \sqrt{\gamma_{b,j}} ||^2 \notag \\
&& \hspace{0.6cm}-\frac{ ||\vect N_{b,s,j} + \vect X_{b,j} \sqrt{\gamma_{b,j}} - \vect \mu_{b,j} ||^2}{\sigbj}\Big ) > \delta \Big ] \\
& = & \Pr  [ \tilde u_b  > F^{-1}_{\tilde{\mathcal X}_1}(1-P_{\text{FA}})  ], \label{eq:136n}
\end{IEEEeqnarray}
where
\begin{equation}
\tilde u_b : = \sum_{j = 1}^q (1-\frac{1}{\sigbj}) ||\vect N_{b,s,j} +\sqrt{\gamma_{b,j}}  \vect X_{b,j} + \frac{1}{\sigbj-1} \vect \mu_{b,j}||^2. 
\end{equation}
Following the same argument provided for the false alarm probability, one can obtain that $\tilde u_b$ is the weighted sum of independent non-central chi-square random variables and thus follows a generalized chi-square distribution as 
\begin{IEEEeqnarray}{rCl}
\tilde u_b &\sim&  \tilde {\mathcal X}_2(\{ w_{b,j}\}_{j =1}^{q}, \ell, \{\tilde \nu_{b,j}\}_{j = 1}^q),
\end{IEEEeqnarray}
where the parameters $\{ w_{b,j}\}_{j =1}^{q}$ and $\{\tilde \nu_{b,j}\}_{j = 1}^q$ are defined in \eqref{eq:27}. 
By denoting $F_{\tilde{\mathcal X}_2} (\cdot)$ as the CDF of the corresponding generalized chi-square distribution, we  conclude the proof. 


\section{Proof of Lemma~\ref{lemma5}}\label{App:E}
To prove this lemma, we follow a similar  argument as the one provided in  \cite[Proposition~1]{MolavianJaziArXiv}. Recall the definition of $\tilde i_{\U} (\vectsf V_b; \vectsf Y_{b,c} )$ from \eqref{eq:iu}. In the following, we prove that  
\begin{IEEEeqnarray}{rCl}
\tilde i_{\U} (\vbj; \ybcj ) \sim \mathcal N \left (\ell \C(\Omega_{b,j}), \ell \mathsf V(\Omega_{b,j})\right) 
\end{IEEEeqnarray}
which consequently proves Lemma~\ref{lemma5}. To this end, recall the definition of $Q_{\ybcj|\vbj} (\vect y_{b,c,j}| \vect v_{b,j})$ and $Q_{\ybcj}(\vect y_{b,c,j})$ from \eqref{eq:qy} and \eqref{eq:qyv}, respectively. Denote 
\begin{equation}
\tilde {\vect Z}_j = \lambda_{b,j} (1-\alc)(\sbjs + (1-\alss)\xbjs) + \nbcj.
\end{equation}
According to the definition of $Q_{\ybcj|\vbj} (\vect y_{b,c,j}| \vect v_{b,j})$,  $\tilde {\vect Z}_j $ follows an i.i.d Gaussian distribution with zero mean and variance $\sigyvj \vectsf I_{\ell \times \ell}$.   
 We thus have
\begin{IEEEeqnarray}{rCl}
\lefteqn{\tilde i_{\U} (\vbj; \ybcj ) = \log \frac{Q_{\ybcj|\vbj}(\vect y_{b,c,j}| \vect v_{b,j})}{Q_{\ybcj}(\vect y_{b,c,j})}} \\
&=& \log \frac{\left (\frac{1}{\sqrt{2\pi \sigyvj}}\right)^{\ell} \exp [-\frac{||\ybcj - \sqrt{\lambda_{b,j}}\vbj||^2}{2\sigyvj}]}{\left (\frac{1}{\sqrt{2\pi \sigy}}\right)^{\ell} \exp [-\frac{||\ybcj ||^2}{2\sigy}]} \\
& = & \ell \C(\Omega_{b,j}) +  \frac{1}{2\sigy}\Big (||\ybcj||^2 -\frac{\sigy}{\sigyvj}||\tilde {\vect Z}_j||^2 \Big) \\
& = &  \ell \C(\Omega_{b,j}) + \frac{1}{2\sigy} \left (\sigy -\sigyvj \right) \left ( \ell -  \frac{||\tilde {\vect Z}_j||^2}{\sigyvj} \right)\notag \\
&& + \frac{\sqrt{\lambda_{b,j}}}{\sigy} \langle \vbj,  \tilde {\vect Z}_j\rangle \\
& = &  \ell \C (\Omega_{b,j})  \notag \\
&& + \frac{1}{2\sigy}\Big (c_1 (\ell - ||\tilde {\vect Z}_j||^2) + c_2 \langle  \tilde {\vect Z}_j, \vbj \rangle \Big)\label{eq:131}
\end{IEEEeqnarray}
where 
\begin{IEEEeqnarray}{rCl}
c_1 &:=& \Omega_{b,j} \sigyvj,\\
c_2 &:=& 2 \sqrt{\lambda_{b,j}}.
\end{IEEEeqnarray}
 The summands in \eqref{eq:131} are not independent, since
  $\vbj$  are not independent across time. One can however express independent uniform random variables on the power shell as a functions of independent Gaussian random variables. To this end,  
 let $\vect W_1 \sim \mathcal N (0, I_{\ell})$ be i.i.d Gaussian random variables independent of the noise $\tilde {\vect Z}_j$. Inputs $V_{b,j,t}$ with $t \in [\ell]$ thus can be expressed as
 \begin{IEEEeqnarray}{rCl}
 V_{b,j,t} & = & \sqrt{\ell (\bc + \alc^2 (1-\bc) )\P}\frac{W_{1,t}}{||\vect W_1||}.
 \end{IEEEeqnarray}
 To apply the CLT for functions proposed in \cite[Proposition~1]{MolavianJaziArXiv}, we consider the sequence $\{\vect U_t : = (U_{1,t}, U_{2,t}, U_{3,t})\}_{t = 1}^{\infty}$ whose elements are defined as 
 \begin{IEEEeqnarray}{rCl}
 U_{1,t} &: =& 1 - \frac{\tilde Z_{j,t}^2}{\sigyvj}, \\
 U_{2,t}&:=&  \sqrt{(\bc + \alc^2 (1-\bc) )\P \sigyvj}W_{1,t} \tilde Z_{j,t},  \\
   U_{3,t}&: =& W_{1,t}^2 -1,  \;
 \end{IEEEeqnarray}
 Note that this random vector has an i.i.d. distribution across time $t = 1, \ldots, n$ and its moments can be easily verified to satisfy $\mathbb E [\vect U_1] = 0$ and $\mathbb E[||\vect U_t||_2^3] < \infty$. 
 The covariance matrix of this vector is given by 
 \begin{equation}
 \text{Cov} (\vect U) = \text{Diag} \left [2, (\bc + \alc^2 (1-\bc) )\P \sigyvj, 2\right ].
 \end{equation}
Next, we define the function $f$ as 
\begin{IEEEeqnarray}{rCl}
f(\vect u) = c_1u_1 + \frac{c_2 u_2}{\sqrt{1+u_3}}. 
\end{IEEEeqnarray}
Notice that $f(\vect 0) = 0$, and all the first and second order partial derivatives of $f$ are continuous in a neighborhood of $\vect u = 0$. The Jacobian matrix $\{\pdv{f(\vect u)}{u_j}\}_{1\times 3}$ at $\vect u = 0$ thus is given by 
\begin{equation}
\mathrm{J} \big |_{\vect u = 0} = [c_1 \; c_2 \; 0 ]. 
\end{equation}
Furthermore,  
\begin{IEEEeqnarray}{rCl}
\lefteqn{f\left (\frac{1}{\ell} \sum_{t = 1}^{\ell} \vect U_t \right) } \notag \\
 &= &\frac{c_1}{\ell} \sum_{t = 1}^{\ell} \left (1-\frac{\tilde Z_{j,t}^2}{\sigyvj}\right)\notag \\
 &&  + \frac{\frac{c_2}{\ell}\sum_{t = 1}^{\ell} \frac{\sqrt{(\bc + \alc^2 (1-\bc) )\P \sigyvj}}{\sqrt{\ell}} W_{1,t} \tilde Z_{j,t}}{\sqrt{1+\frac{1}{\ell} \sum_{t=1}^{\ell}(W_{1,t}^2-1)}}  \notag \\
& = & \frac{1}{\ell} \Big (c_1 (\ell - ||\tilde {\vect Z}_j||^2) + c_2 \langle  \tilde {\vect Z}_j, \vbj \rangle \Big)
\end{IEEEeqnarray}
From the CLT in \cite[Proposition~1]{MolavianJaziArXiv}, we now conclude that the random variable $\tilde i_{\U} (\vbj; \ybcj )$ converges in distribution to a Gaussian distribution with mean $\ell \C(\Omega_{b,j})$ and variance 
\begin{IEEEeqnarray}{rCl}
\lefteqn{\frac{\ell}{4 \sigma_{y,j}^4} [c_1 \; c_2\;   0\; 0 ]  \text{Cov} (\vect U) [c_1 \; c_2\; 0 ]^T } \notag \\
&=& \frac{\ell}{4 \sigma_{y,j}^4} \left [2\Omega_{b,j}^2\sigma_{y|v,j}^4  + 4\lambda_{b,j} (\bc + \alc^2 (1-\bc) )\P \sigyvj  \right]  \notag  \\
& = & \ell \frac{\Omega_{b,j}(2+ \Omega_{b,j})}{2(1+\Omega_{b,j})^2} = \mathsf V(\Omega_{b,j})
\end{IEEEeqnarray}
This concludes the proof.

\end{document}